\documentclass[12pt, draftclsnofoot, onecolumn]{IEEEtran}
\usepackage{cite}
\usepackage{graphicx}
\usepackage[cmex10]{amsmath}
\usepackage{color,soul}
\usepackage{setspace}
\usepackage{amsthm,amssymb}
\usepackage{mathrsfs} 

\newcommand{\squeezeup}{\vspace{-2.5mm}}
\begin{document}
	\title{Game-Theoretic Spectrum Trading in RF Relay-Assisted Free-Space Optical Communications}
	\author{Shenjie Huang, \IEEEmembership{Student Member, IEEE}, Vahid Shah-Mansouri, \IEEEmembership{Member, IEEE}, and Majid Safari, \IEEEmembership{Member, IEEE}\thanks{S. Huang and M. Safari are with the School of Engineering, the University of Edinburgh, Edinburgh (e-mail: \{shenjie.huang, majid.safari\}@ed.ac.uk). V. Shah-Mansouri is with the School of Electrical and Computer Engineering, University of Tehran, Tehran (e-mail: vmansouri@ut.ac.ir).}}
	\maketitle

	\begin{abstract}
		This work proposes a novel hybrid RF/FSO system based on a game theoretic spectrum trading process. It is assumed that no RF spectrum is preallocated to the FSO link and only when the link availability is severely impaired by the infrequent adverse weather conditions, i.e. fog, etc.,  the source can borrow a portion of licensed RF spectrum from one of the surrounding RF nodes. Using the leased spectrum, the source establishes a dual-hop RF/FSO hybrid link to maintain its throughout to the destination. The proposed system is considered to be both spectrum- and power-efficient. A market-equilibrium-based pricing process is proposed for the spectrum trading between the source and RF nodes. Through extensive performance analysis, it is demonstrated that the proposed scheme can significantly improve the average capacity of the system, especially when the surrounding RF nodes are with low traffic loads. In addition, the system benefits from involving more RF nodes into the spectrum trading process by means of diversity, particularly when the surrounding RF nodes have high probability of being in heavy traffic loads. Furthermore, the application of the proposed system in a realistic scenario is presented based on the weather statistics in the city of Edinburgh, UK. It is demonstrated that the proposed system can substantially enhance the link availability towards the carrier-class requirement.       
	\end{abstract}
\section{Introduction}
In recent decades, the scarcity in the radio frequency (RF) spectrum becomes the bottleneck in the expansion of wireless communication networks. As a potential candidate for the long-range wireless connectivity, free-space optical (FSO) communication has attracted widespread and significant interest in both scientific community and industry because of its high achievable data rates, license-free spectrum, outstanding security level and low installation cost. {FSO has numerous applications and in particular it is considered as a cost-effective wireless backhaul solution of the future 5G systems \cite{Dahrouj15}}. However, there exist some limitations and challenges in practical FSO systems including the pointing and misalignment loss due to building sways \cite{Huang17} and unpredictable connectivity in the presence of atmosphere due to the turbulence-induced intensity fluctuation (also known as scintillation) and adverse weather conditions such as rain, snow and fog \cite{survey}.

Beam misalignment fading in terrestrial FSO systems has been accurately modelled \cite{Farid12} and several effective methods have been proposed to mitigate its effects on system performance such as the utilization of beamwidth optimization \cite{Farid:07} and adaptive tracking systems \cite{Bloom:03}. On the other hand, multiple techniques have also been proposed to mitigate the performance degradation caused by scintillation including the spatial diversity at the transmitter \cite{Anguita:07}, at the receiver \cite{Razavi05} or at both transceivers \cite{Navidpour07}, multi-hop relaying \cite{Safari12} and adaptive optics \cite{Tyson:02}. {However, all these mentioned techniques are only useful in the presence of spatially dynamic channel fluctuations.  Adverse weather conditions on the other hand have fairly static characteristics both in time and space, which makes the above techniques ineffective \cite{arnon2012advanced}. Studies have shown that the adverse weather conditions can significantly deteriorate FSO link by introducing an optical power attenuation of up to several hundreds of decibels per kilometer \cite{Kim98}.  Recently, the so-called \textit{hybrid RF/FSO} link has been proposed to effectively improve the link availability of the FSO link by employing an additional RF link \cite{kim2001}.} The motivation behind this idea is that because of the distinct carrier frequencies, FSO links are more susceptible to scattering due to fog and turbulence-induced scintillation whereas RF links are more sensitive to rain conditions (especially for frequencies above $10$ GHz). Therefore, hybrid RF/FSO links can combine the benefits of the two links to combat the effects of adverse weather. 

In the literature, there are basically two main types of hybrid RF/FSO systems based on either switch-over or simultaneous transmission. In switch-over transmission (also called hard-switching transmission) scheme, the RF link is simply a backup link and data is transmitted through either of the channels. 
%It means that only one of the two links is active at a time: when FSO link is in good condition, RF link is idle; otherwise RF link is active and FSO link is idle. 
In \cite{Usman14}, a low-complexity hard-switching hybrid RF/FSO system with both single-threshold and dual-threshold for FSO link operation is proposed. Besides the theoretical studies, several experimental works focusing on this type of hybrid link have also been reported \cite{Lee14}. Although switch-over hybrid RF/FSO link is simple and has also been employed in some commercial FSO products \cite{survey}, the preallocation of RF spectrum to a backup link with occasional use is inherently spectrum-inefficient \cite{Tang}. 

In another type of hybrid RF/FSO links, simultaneous data transmission is considered where both the FSO and RF links are simultaneously active.   One simple implementation of such hybrid links is sending the same data on both channels concurrently and decoding the signal at the receiver based on the more reliable channel  \cite{bloom2002last} or the maximal ratio combining of two channels \cite{Rakia15}. Some other works focus on the designs of joint channel coding and decoding over the two channels in the hybrid link. 
%In \cite{Vangala07}, the hybrid channel coding scheme is proposed using the non-uniform and rate compatible low-density parity check (LDPC) codes.
In particular, the hybrid rateless coding is employed so that the coding rate for each channel in the hybrid link can be adapted to the data rate that the channel can provide and no channel knowledge at the transmitter is required \cite{zhang09,Abdul10}.    
%Later rateless coded hybrid automatic repeat-request (HARQ) scheme for hybrid RF/FSO systems is proposed where instantaneous channel state information is not required and the coding rate is adapted to the data rate that the hybrid link can provide \cite{zhang09,Abdul10}. More recently, the performance of hybrid RF/FSO links is analysed considering the case with and without incremental redundancy HARQ from the information theoretic point-of-view \cite{Makki16}. 
Furthermore, the hybrid RF/FSO systems are also modelled as two independent parallel channels to further improve the total throughout \cite{Jamali16,Najafi17}. 
%For instance, cascaded RF and hybrid RF/FSO system is investigated where multiple users transmit data to the relay node through RF link and the relay forwards the data to the destination through the hybrid RF/FSO link \cite{Jamali16}. 
Although hybrid RF/FSO links with simultaneous transmission outperform those with switch-over transmission, they require both FSO and RF link to be active continuously even when the FSO link is in good conditions and  itself is able to support the required data throughput. Therefore, in the absence of power allocation strategy, hybrid RF/FSO links with simultaneous transmission are power-inefficient and may also generate unneeded RF interference to the environment \cite{Usman14,Rakia15}.

In this work, we propose a novel hybrid RF/FSO system based on the game theoretic spectrum trading. We assume that there exists a preinstalled FSO link between the source and destination, however, no RF spectrum is preallocated to this link. When the link availability is significantly impaired by the infrequent long-term adverse weather conditions, the source attempts to borrow a portion of RF spectrum from one of the surrounding RF nodes, which have licensed spectrum to communicate with the destination, to establish a dual-hop RF/FSO hybrid link and maintain its throughput to the destination. A market-equilibrium-based pricing process is proposed for the spectrum trading between the source and RF nodes. {Compared to above-mentioned hybrid RF/FSO systems in the literature,  the proposed system is considered to be spectrum-efficient since no preallocation of RF spectrum to the link is necessary and the source borrows the RF spectrum only when it is needed. In addition, the investigated system is considered to be power-efficient since the hybrid link is only established during the infrequent adverse weather conditions. Furthermore, the proposed system is also cost-effective by borrowing RF spectrum from surrounding RF nodes rather than establishing and always maintaining a high-cost RF link\footnote{This could be any type of RF link: if sub-$6$ GHz RF link is used the cost of licensing is high; whereas if high-frequency line-of-sight RF link is used the link itself is costly \cite{Dahrouj15}).}}.

%Compared to above-mentioned switch-over hybrid RF/FSO systems, the proposed system is considered to be spectrum-efficient since no RF spectrum preallocation to the link is necessary. On the other hand, compared to the mentioned hybrid RF/FSO systems with simultaneous transmission, the investigated system is considered to be power-efficient since the hybrid link is only established during adverse weather conditions. Furthermore, the proposed system is also cost-effective by borrowing low-frequency RF spectrum from surrounding RF nodes rather than establishing and always maintaining a high-cost RF link (this could be any type of RF link: if low-frequency RF link is used the cost of licensing is high; whereas if high-frequency line-of-sight (LoS) RF link is used the link itself is costly \cite{Dahrouj15}). 

Game theory has been widely employed in the context of wireless networks for resource management especially in cognitive radio networks. For instance, in \cite{Kang} the price-based power allocation strategies for a two-tier femtocell network with a central macrocell underlaid with multiple femtocells is investigated using the Stackelberg game. The frequency spectrum trading between licensed and unlicensed users in the cognitive radio networks is investigated in \cite{Niyato08} where three pricing models including market-equilibrium, competitive and cooperative pricing models are considered. In addition, the game theoretical dynamic spectrum sharing between primary and secondary strategic users is investigated in \cite{Lin11}.   
%In \cite{Lin13}, a hybrid cooperation framework between a mobile operator and a fixed-line operator to provide femtocell services to indoor users is considered using Nash bargaining and sequential game. In addition,  
%we consider a wireless amplify-and-for-ward relay network with one relay node and multiple source-destination pairs/users and propose a pricing framework that enables the relay to set prices to maximize either its revenue or any desirable system utility.  
However, to the best of authors' knowledge, this work is the first time the game-theoretic spectrum trading based hybrid RF/FSO systems are proposed and analysed. 

The rest of this paper is organized as follows. The channel model and system description are shown in Section \ref{System Model}. Section \ref{Spectrum Sharing} presents the derivations of demand and supply functions and describes the spectrum trading scheme and relay selection strategy in detail. The numerical results and discussion are presented in Section \ref{Numerical Result}. Finally, we conclude this paper in Section \ref{conclusion}.     

\section{System Model}\label{System Model}
\begin{figure}[!t]
	\centering
	\includegraphics[width=0.55\textwidth]{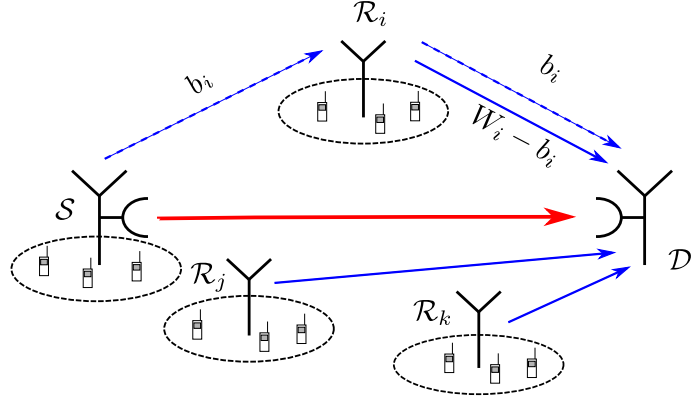}
	\caption{The proposed system model in practical application for wireless backhauling. $\mathcal{S}$: the source; $\mathcal{R}$: the surrounding RF nodes; $\mathcal{D}$: the destination; $b_i$: the leased bandwidth; $W_i$: the total licensed bandwidth the $i$th RF node is allocated for backhauling.  }\label{system}
			\squeezeup
			\squeezeup
\end{figure}
{Fig. \ref{system} shows the schematic of the proposed system for the application of wireless backhauling in a heterogeneous network consists of a large macro-cell and numerous small cells. The source $\mathcal{S}$ denotes the small cell base station (SBS) which requires high data throughput to the macro-cell base station (MBS), i.e., the destination $\mathcal{D}$ and the RF nodes $\mathcal{R}_i$ with $i\in\{1,\ldots,N\}$ are the surrounding SBSs that have wireless backhaul connectivity to $\mathcal{D}$ using licensed sub-$6$ GHz spectrum $W_i$. The source $\mathcal{S}$ would like to send its information to the destination $\mathcal{D}$ and there exists an already installed FSO link between them where a minimal data rate is required. When the data rate of FSO link goes below the required data rate, the source broadcasts a request signal to the surrounding $N$ distributed RF nodes for the sake of buying a portion of their spectrum and establish a hybrid dual-hop RF/FSO  link to improve its data rate to the MBS.}
%There exists an already installed FSO link between the source and destination where a minimal data rate is required. The source keeps monitoring the FSO link condition. When the data rate of FSO link goes the required data rate, the source broadcasts a signal to the surrounding $N$  distributed RF nodes $\mathcal{R}_i$ with $i\in\{1,\ldots,N\}$, which have licensed RF spectrum to communicate with the destination, for the sake of buying a portion of their spectrum to improve its data rate to the destination. The proposed novel communication setup can model various practical applications. \textcolor{blue}{One possible application is the wireless backhauling in a heterogeneous network with a large macro-cell in combination with numerous small cells as presented in Fig. \ref{system} (b) \cite{Li15}. In this application, the source denotes the small cell base station (SBS) which requires high data throughput to the macro-cell base station (MBS), i.e., the destination, and RF nodes are the surrounding SBSs that have wireless backhaul connectivity to the final destination whereas with lower throughput requirement using licensed sub-$6$ GHz spectrum. 
{Compared to the line-of-sight (LoS) RF backhauling with unlicensed high-frequency spectrum (e.g., millimeter-wave), the RF backhauling with low-frequency licensed spectrum is also widely investigated  \cite{Jamali16,Najafi17,zhang09} due to its advantage of non-LoS property and lower installation cost, which makes it more attractive on small-cell networks especially in urban areas \cite{Dahrouj15,Siddique15}.
The proposed novel communication scheme can also be employed in other applications where the surrounding RF nodes are any types of relay nodes in LTE-based wireless backhaul architecture \cite{Wang15}.}

\subsection{Channel Model}\label{CM}
Since both FSO and RF links are involved in the proposed system, the channel models for both channels need to be investigated. 

\subsubsection{\textbf{FSO link}} Considering that adaptive tracking systems are employed to properly address the misalignment fading, the FSO link  $\mathcal{S}-\mathcal{D}$ suffers from two main channel impairments including the turbulence-induced scintillation and adverse weather condition whereas at very different time scales. The scintillation is a short-term effect with coherence time $T_s$ on the order of several milliseconds \cite{Bloom:03}, however, the weather condition is a long-term effect with time-scale $T_{c}$ on the order of hours \cite{zhang09}. Assuming that the FSO link employs the intensity modulation direct detection (IM/DD), the channel expression can be written as
\begin{equation}\label{FSO_chan}
s_{o}=\rho g_{o} h_{o}x_{o}+z_{o},
\end{equation} 
where $\rho$ is the the responsivity of the photodetector, $g_{o}$ refers to the average power gain, $h_{o}$ denotes the random turbulence-induced intensity fading, $x_{o}$ is the transmitted optical intensity, $s_{o}$ is the received electrical signal and $z_{o}$ is zero-mean real Gaussian noise with variance $\sigma_{o}^2$. Note that we use the subscript `$o$' to denote the optical link. The signal-independent Gaussian noise $z_{o}$ in (\ref{FSO_chan}) arises from thermal noise as well as the shot noise induced by the ambient light.
%which is considered to be independent of the signal itself. 
%Shot noise (also called the quantum noise) is generated from the random fluctuation of the current flowing through the PD and can be modelled by a Poisson distribution \cite{survey}. With relatively high received optical intensity, shot noise can be approximated as a Gaussian random variable with high accuracy\cite{gagliardi}. Note that in this work the signal-induced shot noise which is inherently signal-dependent is considered to be negligible compared to the shot noise caused by background radiation and thermal noise. 
The average gain $g_{o}$ can be expressed as \cite{he09,Jamali16}
\begin{equation}\label{hfa}
g_{o}=\left[\mathrm{erf}\left(\frac{\sqrt{\pi}d}{2\sqrt{2}\phi L_{SD}}\right)\right]^2\times e^{-\kappa L_{SD}},
\end{equation} 
where the first and second term denote the geometric loss due to the divergence of the transmitted beam and weather-related atmospheric loss due to scattering and absorption, respectively, $d$ is the receiver aperture diameter, $\phi$ is the beam divergence angle, $L_{SD}$ is the distance between the source and the destination, and $\kappa$ is a weather-dependent attenuation coefficient determined based on the Beer-Lambert law. 
%\textcolor{blue}{According to the Beer-Lambert law, the relationship between $\kappa$ and visibility $V$ in km can be expressed as $10^{-\kappa L_{SD}/10}=e^{-\psi L_{SD}}$   
%given by $\kappa=0.01\psi\,\mathrm{log}_{10}e$ 
The relationship between $\kappa$ and visibility $V$ in km can be expressed as \cite{Kim98}
\begin{equation}\label{visi}
\kappa=\frac{3.91}{V}\left(\frac{\lambda_{o}}{550\times10^{-9}}\right)^{-\zeta},
\end{equation}  
where $\lambda_{o}$ is optical wavelength and $\zeta$ is the size distribution of the scattering particles equal to $1.6$, $1.3$ and  $0.585V^{1/3}$ when $V>50$, $6\leq V\leq 50$ and $V<6$, respectively. 
%\begin{equation}\label{zeta}
%\zeta=\left\{
%\begin{array}{ll}
%1.6, &V>50,\\
%1.3, &6\leq V\leq 50,\\
%0.585V^{1/3}, & V<6.
%\end{array}
%\right.
%\end{equation} 
There are several ways to model the turbulence-induced intensity fluctuation $h_{o}$ such as log-normal distribution and Gamma-Gamma distribution. In this work, we employ the Gamma-Gamma distribution which can describe the intensity function within a wide range of turbulence conditions as \cite{he09}
\begin{equation}
f_{h_{o}}(x)=\frac{2\left(\alpha \beta \right)^{\left(\alpha+\beta\right)/2}}{\Gamma(\alpha)\Gamma(\beta)}{x}^{\left(\alpha+\beta\right)/2-1} K_{\alpha-\beta}\left(2\sqrt{\alpha\beta x}\right),
\end{equation}
where $\Gamma(\cdot)$ is the Gamma function, $K_p(\cdot)$ is the modified Bessel function of the second kind. The parameter $\alpha$ and $\beta$ are given by
\begin{equation}
\alpha\!=\!\!\bigg[\mathrm{exp}\!\bigg(\! \frac{0.49\chi^2}{\left(1\!+\!0.18\vartheta^2\!+\!0.56\chi^{12/5}\right)^{7/6}}\bigg)\!-1\!\bigg]^{-1}\!\!,
\beta\!=\!\!\bigg[\mathrm{exp}\!\bigg(\! \frac{0.51\chi^2\left(1\!+\!0.69\chi^{12/5}\right)^{-5/6}}{\left(1\!+\!0.9\vartheta^2+\!0.62\vartheta^2\chi^{12/5}\right)^{5/6}}\bigg)\!-1\!\bigg]^{-1}\!\!,
\end{equation}
respectively, where $\chi^2=0.5C_n^2k^{7/6}L_{SD}^{11/6}$, $\vartheta^2=kd^2/4L_{SD}$ and $k=2\pi/\lambda_{o}$. Note that $C_n^2$ is the turbulence refraction structure parameter. For IM/DD FSO channel given in (\ref{FSO_chan}), the achievable rate (channel capacity lower bound) in the presence of average transmitted optical power constraint, i.e., $E[x_{o}]\leq P_{o}$, conditioned on the random channel gain $h_{o}$ can be expressed as \cite{Zedini16}
\begin{equation}\label{CFSO_ins}
C_{o}=\frac{W_{o}}{2}\mathrm{log}_2\left(1+\frac{e\rho^2h_{o}^2P_{o}^2}{2\pi\sigma^2_{o}}\right),
\end{equation}     
where $W_{o}$ is the bandwidth of the FSO link. Note that different from the traditional AWGN channel capacity expression in RF, the SNR term in (\ref{CFSO_ins}) is proportional to the squared optical power due to the employed intensity modulation. 

In the proposed system, setting a reasonable performance metric to decide whether the FSO link is satisfactory or not is crucial for defining the trigger of switching between FSO-only link and hybrid RF/FSO link. {In this work, we employ the sliding window averaging strategy with a relatively long window interval compared to the scintillation coherence time $T_s$ to smooth out the quick FSO link capacity fluctuations introduced by scintillation, therefore the measured average link capacity can accurately reflect the current long-term weather condition \cite{Izadpanah03}.}
%Since the time scale of weather change is much longer than that of scintillation, i.e., $T_c \gg T_s$, one possible choice of FSO link performance metric could be the real-time FSO link capacity on the condition of the short-term instantaneous scintillation. However, choosing this performance metric unavoidably results in high-frequency switching between two link types on every $T_s$ interval, which is not practically suitable. 
%In this work, we employ the sliding window averaging strategy to smooth out the quick FSO link capacity fluctuation due to the short-term fading as in \cite{Izadpanah03} so that the FSO performance metric can accurately reflect the long-term weather conditions. We consider a simple rectangular window shape with a relatively long window interval $T_w$ compared with the scintillation coherence time, i.e., $T_w\gg T_s$. \textcolor{blue}{The time scales considered in this work is summarised in Fig. \ref{time_scales}.} 
{The measured average FSO link capacity over the window interval is selected as the performance metric which is approximated as} $\overline{C}_{o}={E}\left[C_{o}\right]$ where ${E}[\cdot]$ denotes the ensemble expectation. The source keeps monitoring the FSO link condition and calculate $\overline{C}_{o}$ every window interval. When $\overline{C}_{o}$ is lower than the minimal data rate requirement $C_{th}$, the spectrum trading process is triggered to establish the dual-hop RF relay link, whereas when it exceeds $C_{th}$, the source will stop buying the RF spectrum. Note that to ensure the source could make quick response to the change of weather conditions, the window interval should be set much less than the time-scale of the weather changes $T_c$. {Also note that $C_{th}$ can be any positive values and larger $C_{th}$ indicates higher data rate requirement, which results in more frequent spectrum trading events between the relays and source and also higher system complexity.}    

\subsubsection{\textbf{RF link}} 
In this work, it is assumed that the source does not establish and maintain a direct RF link along the FSO link to the destination, but instead it uses the borrowed spectrum from surrounding RF nodes to relay its data to the destination only when needed. 
%
%In this work, it is assumed that there is no direct RF link between the source and destination, i.e., the source can use the leased RF spectrum to send its data to destination only via the surrounding RF nodes. This assumption is considered to be valid for example when a large separation and weak RF transmitted power are employed \cite{Rankov07} or when the data transmission from the source is designed solely based on the channel from the source to the RF relay node \cite{Zhang092}. 
When a RF node $\mathcal{R}_i$ is selected by the source as the relay to realize the dual-hop  $\mathcal{S}-\mathcal{R}_i-\mathcal{D}$ link, this RF relay channel can be expressed as
\begin{equation}\label{RF_chan}
s_{R,i}^{(t)}=\sqrt{g_{R,i}^{(t)}}h_{R,i}^{(t)}x_{R}^{(t)}+z_{R,i}^{(t)},
\end{equation}
where the subscript `$R$' is used to denote the RF link,  $t=\{1,2\}$ refer to the RF link from the source to the relay ($\mathcal{S}-\mathcal{R}_i$ link) and that from the relay to the destination ($\mathcal{R}_i-\mathcal{D}$ link), respectively, $s_{R,i}^{(t)}$ and $x_{R}^{(t)}$ are the received and transmitted signal, respectively, $g_{R,i}^{(t)}$ denotes the average power gain, $h_{R,i}^{(t)}$ is the RF fading coefficient and $z_{R,i}^{(t)}$ refers to the zero-mean complex Gaussian noise with power spectrum density $N_0$. For RF signal transmission it is considered that Gaussian codebooks are employed at transmitters, which means at every symbol duration the transmitted symbol $x_{R}^{(t)}$ is generated independently based on a zero-mean rotationally invariant complex Gaussian distribution \cite{Jamali16}. The RF transmitted power, i.e., $E[|x_{R}^{(t)}|^2]$, at the source and the relay are denoted as $P_{R}^\mathcal{S}$ and $P_{R}^\mathcal{R}$, respectively. 
%In this paper we assume that these two transmitted powers are equal for simplicity, i.e., $P_{R}^\mathcal{S}=P_{R}^\mathcal{R}=P_{R}$. 
%Furthermore, it is worth noting that in (\ref{RF_chan}) the RF link is modelled as a fading-free additive white Gaussian noise channel. This is a reasonable assumption considering that in most commercial RF systems, sufficient fading margin is employed in the link budget to ensure high reliability and in addition numerous effective diversity schemes can be applied to mitigate the fading effects effectively \cite{zhang09}. 
The average power gain $g_{R,i}^{(t)}$ can be expressed as \cite{Jamali16}
\begin{equation}
g_{R,i}^{(t)}=\left[\frac{\sqrt{G_{TX}G_{RX}}\lambda_{R}}{4\pi L_\mathrm{ref}}\right]^2\times \left(\frac{L_\mathrm{ref}}{L_{R,i}^{(t)}}\right)^\delta,
\end{equation}
where $G_{TX}$ and $G_{RX}$ refer to the RF transmitter and receiver gain, respectively, $\lambda_{R}$ denotes the RF wavelength, $L_\mathrm{ref}$ is the reference distance for the antenna far-field, $L_{R,i}^{(t)}$ denotes the link distance with $t=1,2$ for $\mathcal{S}-\mathcal{R}_i$ and $\mathcal{R}_i-\mathcal{D}$ link, respectively, and $\delta$ refers to the RF path-loss exponent. Note that the average power gain $g_{R,i}^{(t)}$ is associated with the specific distance between the source (relay) and relay (destination) and the fading coefficient $h_{R,i}^{(t)}$ can be described by distinct models such as Rayleigh, Rician or Nakagami-m fading in different application scenarios \cite{Dahrouj15,Rakia15,Ahumada05}.      
%for the fading coefficient $h_{R,i}^{(t)}$, we model its amplitude as Rayleigh distribution with PDF
%\begin{equation}
%f_{|h_{R,i}^{(t)}|}(u)=\frac{2u}{\Omega}\mathrm{exp}\left(\frac{-{u}^2}{\Omega}\right),
%\end{equation}
%where parameter $\Omega$ is the power of the fading and after normalization one has $\Omega=1$. In this paper, by employing Rayleigh fading model, it is assumed that there is no LoS path between the the source (relay)  and the relay (destination). This is a reasonable assumption when the investigated system is located at environments where the LoS condition is hard to be guaranteed, e.g., urban environment with large buildings \cite{Dahrouj15,Siddique15}. However, we would like to emphasize that other RF multipath fading models such as Rician fading or Nakagami-m fading might also be employed under some other specific scenarios \cite{Rakia15,Ahumada05}. 

According to the channel expression of RF links given in (\ref{RF_chan}), the channel capacity for the $\mathcal{S}-\mathcal{R}_i$ link and $\mathcal{R}_i-\mathcal{D}$ link can be expressed as
\begin{equation}\label{Crf}
C_{R,i}^{(1)}(b_i)=b_i\,\mathrm{log}_2\left(1+\frac{g_{R,i}^{(1)}|h_{R,i}^{(1)}|^2P_{R}^\mathcal{S}}{N_0b_i}\right), \quad  C_{R,i}^{(2)}(b_i)=b_i\,\mathrm{log}_2\left(1+\frac{g_{R,i}^{(2)}|h_{R,i}^{(2)}|^2P_{R}^\mathcal{R}}{N_0W_i}\right),  
\end{equation}
respectively, where $b_i$ is the size of spectrum borrowed from the relay and $W_i$ is the total licensed spectrum that the $i$th relay possesses for $\mathcal{R}_i-\mathcal{D}$ link. From (\ref{Crf}) one can also see that the spectral efficiency of $\mathcal{S}-\mathcal{R}_i$ link is related to the size of the leased spectrum $b_i$. This is because the transmitted RF power is assumed to be fixed, hence increasing $b_i$ will introduce more noise and reduce the received SNR which results in the decrease of the spectral efficiency \cite{Lin13}. On the other hand, the spectral efficiency of the $\mathcal{R}_i-\mathcal{D}$ link is not associated with the size of leased bandwidth $b_i$, since the signal-to-noise ratio (SNR) of the $\mathcal{R}_i-\mathcal{D}$ link is determined by the fixed total transmitted power and total bandwidth. With the expressions of the RF link capacity given in (\ref{Crf}), the capacity of the decode-and-forward RF link  $\mathcal{S}-\mathcal{R}_i-\mathcal{D}$ can be written as \cite{Siddique15}
\begin{equation}\label{Ci}
C_{R,i}(b_i)=\mathrm{min}\left\{q_i\,C_{R,i}^{(1)}(b_i)\,, (1-q_i)\,C_{R,i}^{(2)}(b_i)\right\},
\end{equation}
where $q_i\in(0,1)$ refers to a time sharing variable. We use $q_i$ to indicate the fraction of time when $\mathcal{S}-\mathcal{R}_i$ link is active and hence $1-q_i$ denotes the time fraction when $\mathcal{R}_i-\mathcal{D}$ link is active.\footnote{It is worth noting that the RF node continuously transmits its own backhaul data using the remaining $W_i-b_i$ spectrum.}  The capacity of the RF relay link can be expressed as in (\ref{Ci}) on the condition that the RF relay is operated on a half-duplex mode \cite{Jamali16,Najafi17}. 
%which means that it can either receive from the source or transmit to the destination over the RF link with the shared bandwidth and cannot do both simultaneously \cite{Jamali16}. We would like to emphasize that full-duplex RF nodes \cite{Siddique15} can also be involved in the system,  however, due to the broadcast nature of RF signals, high hardware complexity for efficient self-interference suppression is required in full-duplex communication. Therefore in this work we focus on the half-duplex transmission for RF nodes because of its simplicity and feasibility \cite{Najafi17}. 
Since the first term in the min function of (\ref{Ci}) is a monotonically increasing function of $q_i$ and the second term is a monotonically decreasing function of $q_i$, for a given shared spectrum $b_i$ the optimal $q^*_i$ to maximize  $C_{R,i}$ is given by the cross point of the two functions as  
\begin{equation}\label{opt_p}
q^*_i=\frac{r_i}{y(b_i)+r_i}.
\end{equation}
where $y(b_i)$ and $r_i$ is the spectral efficiency of the 
$\mathcal{S}-\mathcal{R}_i$ link and $\mathcal{R}_i-\mathcal{D}$ link given by 
\begin{equation}\label{1der}
y(b_i)=\mathrm{log}_2\left(1+\frac{v_i}{b_i}\right),\, r_i=\mathrm{log}_2\bigg(1+\frac{g_{R,i}^{(2)}|h_{R,i}^{(2)}|^2P_{R}^\mathcal{R}}{N_0W_i}\bigg),  
\end{equation}
respectively, with $v_i={g_{R,i}^{(1)}|h_{R,i}^{(1)}|^2P_{R}^\mathcal{S}}/{N_0}$.
This optimal time sharing variable (\ref{opt_p}) indicates that the capacities of $\mathcal{S}-\mathcal{R}_i$ and  $\mathcal{R}_i-\mathcal{D}$ links are identical, which means all the data transmitted to the relay can be successfully transferred to the destination. By substituting (\ref{opt_p}) into (\ref{Ci}) one can hence get the maximal capacity of the dual-hop $\mathcal{S}-\mathcal{R}_i-\mathcal{D}$ link as
\begin{equation}\label{CRF}
C_{R,i}(b_i)=\frac{ b_i\,r_i\,y(b_i)  }{y(b_i)+r_i}.
\end{equation}
To check the monotonicity of $C_{R,i}(b_i)$, one can take the first derivative of it as
\begin{equation}\label{Tbinew}
\mathcal{T}(b_i)=\frac{\mathrm{d} C_{R,i}(b_i)}{\mathrm{d} b_i}=\frac{r_i\left[y(b_i)\right]^2+r_i^2y(b_i)-\frac{r_i^2}{\mathrm{ln}2}\left[1-2^{-y(b_i)}\right]}{\left[y(b_i)+r_i\right]^2}.
\end{equation}
It can be proved that $\mathcal{T}(b_i)>0$ holds for all $b_i\geq 0$ and approaches to zero for $b_i\to +\infty$, thus $C_{R,i}(b_i)$ is a monotonically increasing function of $b_i$ with a saturated value $v_i/\mathrm{ln}2$ at high $b_i$.

\subsection{{Spectrum Trading Game}} \label{System Description}
We consider that each RF node has data traffic and hence has its own backhaul data to transmit to the destination using a maximum allocated bandwidth $W_i$ via the $\mathcal{R}_i-\mathcal{D}$ link. Particularly, in wireless backhaul application presented in Fig. \ref{system} this data traffic comes from the connected user equipments (UEs) in the small cell. In addition, it is assumed that different RF nodes are allocated non-overlapping frequency spectrum for data transmission so that there is no interference between them at the destination \cite{Zhang08}. {When RF nodes are not in high traffic load, they might be willing to lend part of their spectrum to the source and obtain some revenues at the expense of self-data transmission limitation. In the wireless backhaul application, this limitation may effect the QoS performance of the connected UEs in the small cells.}

When the RF nodes notice that the source requests to buy their spectrum due to its non-functional FSO link condition, a two-player game will be operated between the source and each RF nodes. In this paper, we consider a market-equilibrium-based pricing approach for the spectrum trading game \cite{Niyato08,Niyato082,Tehrani12} where the source is treated as the \textit{buyer} and the RF relay nodes are treated as the \textit{sellers}. It is assumed that different RF nodes are not aware of each other\footnote{{In practice, this assumption is justified due to the lack of any centralized controller or information exchange among RF nodes. When the RF nodes have more information about each other, more complicated spectrum trading games can be investigated such as competitive and cooperative games \cite{Niyato08,Niyato083}.}}  		
%However, if the RF nodes are aware of the existence of others, they would compete with each other to sell their spectrum, which results in a game with competitive pricing \cite{Niyato08}. Furthermore, if extensive communication among RF nodes is possible, they could even cooperate to obtain highest total profit by lending spectrum to the source, which leads to the cooperative pricing \cite{Niyato083}.
and each seller has to negotiate with the source and sets their price independently to meet the buyer's demand according to their own utilities. In the spectrum trading games (discussed later in Section \ref{Spectrum Sharing}), both source and RF nodes in the system are considered to be rational and selfish so that they only focus on their own payoff and always follow the best strategies which maximize their own utility. When all games (or negotiations) are finished, the source will receive the distinct unit spectrum prices proposed by different RF nodes and the sizes of spectrum that they want to lend. Based on this received information, the source is able to decide on the best RF node. 
\begin{figure}[!t]
	\centering
	\includegraphics[width=0.56\textwidth]{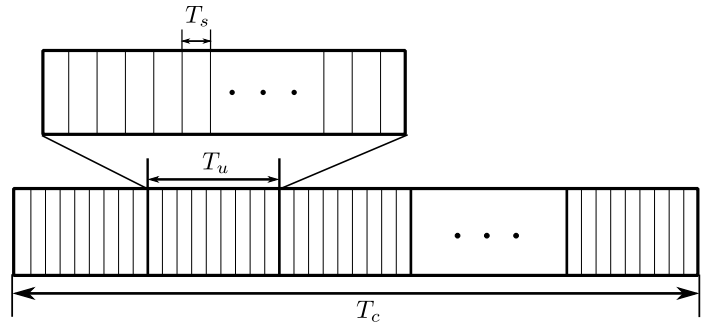}
	\caption{Time scales: $T_s$, the coherence time of the scintillation; $T_c$, the coherence time of the weather condition; $T_u$, time duration of using the leased RF spectrum before repeating the game.}\label{time_scales}
	\squeezeup
	\squeezeup
\end{figure}
By notifying the selected RF node, a dual-hop RF relay link $\mathcal{S}-\mathcal{R}_i-\mathcal{D}$ can be established and the previous FSO link $\mathcal{S}-\mathcal{D}$ turns to a hybrid dual-hop RF/FSO link which is designed to provide a data rate higher than the required data rate. {The proposed system with single RF relay selection benefits from its simplicity. However, it is also possible for the source to borrow RF spectrum from multiple RF nodes simultaneously rather than routing its traffic from a single node (similar to the system investigated in \cite{Wang09} in the context of RF relaying system), which can be an interesting work to address in the future.}  

{The result of the spectrum trading game mainly depends on the condition of the RF relay link and the traffic load supported by surrounding RF nodes. As a result, the source should repeat the relay selection when these conditions significantly change. Denote the coherence time of the fading in RF link and traffic loads in surrounding nodes as $T_f$ and $T_t$, respectively. For wireless RF links with fixed transceivers the temporal behaviour of the fading is more stable compared to the mobile channels and is tightly related to the existence of LoS and environmental conditions (e.g., the vehicular traffic). It is concluded that $T_f$ in the fixed wireless RF links is typically on the order of seconds \cite{Ahumada05,Gans02}. On the other hand, $T_t$ is relatively longer which depends on different application scenarios and could vary from several seconds to hours. Therefore, it is reasonable to assume that the source uses the leased RF spectrum for a time period of $T_u=\mathrm{min}\{T_f, T_t\}$ before involving into a new spectrum trading game and possibly updating the selected RF relay. The time scales considered in this work are plotted in Fig. \ref{time_scales}. Invoking the discussion of the coherence time of scintillation $T_s$ and weather change $T_c$ in Section \ref{CM}, one can get that $T_u$ is much shorter than $T_c$ but much longer than $T_s$. Therefore, in every $T_c$ interval the source should repeat the RF relay selection based on the updated information every $T_u$ interval.}

\section{Solution of Spectrum Trading} \label{Spectrum Sharing}
In this section, we present the solution of the game-theoretic spectrum trading process for the proposed communication setup, which is triggered under the condition $\overline{C}_{o}<C_{th}$. 
%One reasonable choice of $C_{th}$ could be the outage threshold of the FSO link \cite{Farid:07}, which means that the spectrum sharing process is only triggered when the FSO link is in outage. In this work, we consider that $C_{th}$ equals to the outage threshold. 
%In terms of the game-theoretic spectrum sharing process, 
we consider a market-equilibrium-based pricing model in which, for a given unit spectrum price, the buyer (source) chooses its spectrum demand based on its \textit{demand function} and the sellers (the surrounding RF nodes) set the amount of spectrum they would like to offer according to their \textit{supply function}. The market-equilibrium refers to the price in which the spectrum demand of the source equals to the spectrum supply of the relays and no excess supply exists in the market \cite{Tehrani12,Niyato082}. {In the proposed system, the same spectrum trading process (game) should be applied between the source and each surrounding RF node. Without loss of generality, we will firstly focus on the spectrum trading between the source $\mathcal{S}$ and the $i$th RF nodes $\mathcal{R}_i$.} Later in Section \ref{relay selection}, the relay selection issue will be discussed.         
\subsection{Utility of Source and the Demand Function}\label{demandfunc}
To quantify the spectrum demand of the source when choosing $\mathcal{R}_i$ as the relay node, the utility gained by the source should be determined which can be expressed as \cite{Lin13} 
\begin{equation}\label{uti_sour}
\mathcal{U}_i=\lambda C_{R,i}(b_i)-b_ip_i,
\end{equation}
where $\lambda$ is the constant weight indicating the obtained revenue per unit transmission rate and $p_i\in(0,+\infty)$ is the unit spectrum price offered by the RF node $\mathcal{R}_i$. Equation (\ref{uti_sour}) indicates that the source gains revenue from the data rate improvement by borrowing spectrum from the relay at the cost of paying the leased spectrum. By substituting (\ref{CRF}) into (\ref{uti_sour}), the utility function can be rewritten as
\begin{equation}\label{uti_1}
\mathcal{U}_i=\frac{\lambda b_i\,r_i\,y(b_i)  }{y(b_i)+r_i}-b_ip_i. 
\end{equation}
The so-called demand function refers to the spectrum demand that maximizes the utility function (\ref{uti_1}) when the spectrum price $p_i$ is given \cite{Niyato082}. {To establish a RF relay link  $\mathcal{S}-\mathcal{R}_i-\mathcal{D}$, the optimization problem at $\mathcal{S}$ can be expressed as follows:} 
\begin{align}\label{opti1}
\underset{b_i}{\mathrm{max}} \;\;&\mathcal{U}_i= \frac{\lambda b_i\,r_i\,y(b_i)  }{y(b_i)+r_i}-b_ip_i,\\\nonumber
\mathrm{s.t.}\;\; & C_{R,i}(b_i)\geq C_{th}-\overline{C}_{o}, \quad b_i> 0, \quad \mathcal{U}_i>0,
\end{align}
%where $C_{ex}$ denotes the expected data rate between the source and destination after establishing the dual-hop RF/FSO hybrid link and $E_{T_u}[{C}_{o}]$ refers to the average FSO link capacity over the $T_u$ interval in which the spectrum trading results will remain. Based on the discussion of the coherence times in Section \ref{System Model}, it can be assumed that the time-scale of FSO intensity fluctuation $T_s$ is much less than $T_u$, therefore $E_{T_u}[{C}_{o}]$ can be approximated as the ensemble expectation $\overline{C}_{o}$ given in (\ref{ensemble}).       
{The first condition in (\ref{opti1}) ensures that the capacity of the RF dual-hop relay $\mathcal{S}-\mathcal{R}_i-\mathcal{D}$ link should be larger than the $C_{th}-\overline{C}_{o}$ so that by establishing the hybrid link, the total data rate between the source and destination, i.e., $C_{R,i}(b_i)+\overline{C}_{o}$, is higher than the minimum data rate requirement $C_{th}$. 
%Note that this condition can also be removed if the source put no data rate requirement on the established relay link. 
%In this work, we assume that this expected data rate is equal to $C_{th}$, i.e., $C_{ex}=C_{th}$, which indicates that through spectrum sharing the source expects to get a data rate higher than the minimal data rate requirement. 
The second condition in (\ref{opti1}) indicates that the leased bandwidth is positive and the last condition guarantees that the source can achieve positive utility. 
%Note that the proposed game between the source and RF node is only triggered when the average capacity of FSO link does not achieve the required data rate, hence, if the $\mathcal{S}-\mathcal{R}_i-\mathcal{D}$ link can be successfully established, the source has to borrow positive bandwidth, i.e., $b_i>0$, from the relay to obtain additional data rate. 
%This condition is put since when the utility is non-positive, which means that the benefit of borrowing the spectrum from the relay node is less than its cost, the source will quit the trading. 
{The solution of the optimization problem (\ref{opti1}) gives the optimal spectrum demand denoted as $\mathscr{D}_i(p_i)$ as a function of the price $p_i$. }} 

\newtheorem{Proposition}{Proposition}
{\begin{Proposition}
The solution to the optimization problem (\ref{opti1}) can be expressed as
\begin{equation}\label{demand}
\mathscr{D}_i(p_i)=\left\{
\begin{array}{ll}
b^\mathrm{root}_i &0< p_i<\lambda \mathcal{T}(b_i^\mathrm{min}),\\
b_i^\mathrm{min} &\lambda \mathcal{T}(b_i^\mathrm{min})\leq p_i< \frac{\lambda r_iy(b_i^\mathrm{min})}{y(b_i^\mathrm{min})+r_i},\\
0 & p_i\geq \frac{\lambda r_iy(b_i^\mathrm{min})}{y(b_i^\mathrm{min})+r_i},
\end{array}
\right.
\end{equation} 
where $b_i^\mathrm{min}$ denotes the positive root of the equation  
\begin{equation}\label{bmin}
(C_{th}-\overline{C}_{o})\left[y(b_i)+r_i\right]=b_i\,r_i \,y(b_i), 
\end{equation}
and $b^\mathrm{root}_i$ refers to the positive root of the equation 
\begin{equation}\label{first_dri3}
\mathcal{T}(b_i)=p_i/\lambda,
\end{equation}  
when $C_{th}-\overline{C}_{o}\leq v_i/\mathrm{ln}2$ is satisfied.  However when  $C_{th}-\overline{C}_{o}> v_i/\mathrm{ln}2$ holds, the source will quit the spectrum trading game resulting in zero spectrum demand, i.e., $\mathscr{D}_i(p_i)=0$,
\end{Proposition}}

\begin{proof}
To solve the optimization problem (\ref{opti1}), we should firstly check the convexity of the objective function. The second derivative of (\ref{uti_1}) with respect to $b_i$ can be expressed as
\begin{equation}\label{2der}
\frac{\mathrm{d}^2 \mathcal{U}_i}{\mathrm{d} b_i^2}=\lambda \frac{\mathrm{d}^2 C_{R,i}(b_i)}{\mathrm{d} b_i^2}=\frac{-\lambda r_i^2v_i^2}{\mathrm{ln}2\,b_i\left(y(b_i)+r_i\right)^3\left(b_i+v_i\right)^2}\left[y(b_i)+r_i+\frac{2}{\mathrm{ln}2}\right],
\end{equation}
One can see that for $b_i>0$, $\frac{\mathrm{d}^2 \mathcal{U}_i}{\mathrm{d} b_i^2}<0$ always holds, therefore the objective function in (\ref{opti1}) is concave. The optimal $b_i$ can therefore be calculated through the KKT conditions. The Lagrangian associate with this optimization problem can be written as
\begin{equation}
\mathscr{L}=-\lambda C_{R,i}(b_i)+\gamma_1\left[C_{th}-\overline{C}_{o}-C_{R,i}(b_i)\right]-\gamma_2b_i+b_ip_i-\gamma_3\left[\lambda C_{R,i}(b_i)-b_ip_i\right],
\end{equation}
and the corresponding KKT conditions can be expressed as
\begin{align}\label{KKT1}
&b_i>0,\,\, \,C_{R,i}(b_i)\geq C_{th}-\overline{C}_{o},\,\,\, \lambda C_{R,i}(b_i)>b_ip_i,\, \,\,\gamma_1,\gamma_2,\gamma_3\geq 0, \\\nonumber
&\gamma_2b_i=0, \, \gamma_3\left[\lambda C_{R,i}(b_i)-b_ip_i\right]=0,\,\,{\mathrm{d} \mathscr{L}}/{\mathrm{d} {b_i}}=0, \,\, 
\gamma_1\left[C_{th}-\overline{C}_{o}-C_{R,i}(b_i)\right]=0.
\end{align}
To solve the KKT conditions (\ref{KKT1}), let's firstly focus on the given conditions except the last two equalities. Since the capacity $C_{R,i}(b_i)$ given in (\ref{CRF}) is a monotonically increasing function of $b_i$, the condition $C_{R,i}(b_i)\geq C_{th}-\overline{C}_{o} $ indicates that $b_i\geq b_i^\mathrm{min}$ where $b^\mathrm{min}_i$ is given by the root of the nonlinear equation (\ref{bmin}). It can be proved that with the constraint $C_{th}-\overline{C}_{o}\leq v_i/\mathrm{ln}2$, a single positive root for the nonlinear equation (\ref{bmin}) denoting $b_i^\mathrm{min}$ always exists, which can be calculated numerically. The bandwidth $b_i^\mathrm{min}$ hence refers to the minimum bandwidth that the source requests from the relay $\mathcal{R}_i$. On the other hand, if $C_{th}-\overline{C}_{o}> v_i/\mathrm{ln}2$ holds, there is no positive root for (\ref{bmin}). It means that the source requests a data rate too high so that the RF relay link cannot provide even when infinite bandwidth is leased. In this scenario, the source will not borrow spectrum from the relay and quit the game with $\mathscr{D}_i(p_i)=0$. 

When the condition $b_i\geq b_i^\mathrm{min}$ is met, the condition $\gamma_2b_i=0$ necessitates $\gamma_2=0$. In addition, the equality $\gamma_3\left[\lambda C_{R,i}(b_i)-b_ip_i\right]=0$ and inequality $\lambda C_{R,i}(b_i)>b_ip_i$ together indicates $\gamma_3=0$. One can further calculate that the inequality $\lambda C_{R,i}(b_i)>b_ip_i$ necessitates $y(b_i)>p_ir_i/(\lambda r_i-p_i)$ when $p_i<\lambda r_i$. Note that considering the definition of the source utility given in (\ref{uti_1}), if $p_i\geq\lambda r_i$, one has $\mathcal{U}_i<0$ which against our condition of positive utility. Therefore, the constraint on the unit price $p_i<\lambda r_i$ has to be satisfied. Based on the conditions discussed above, we have the constraint for $b_i$ given by 
\begin{equation}\label{rangeb}
b_i^\mathrm{min}\leq b_i<y^{-1}\left(\frac{p_ir_i}{\lambda r_i-p_i}\right),
\end{equation}
with the condition on $p_i$ that $p_i<\lambda r_i$ where $y^{-1}(\cdot)$ refers to the inverse function of $y(\cdot)$ given in (\ref{1der}). To further justify (\ref{rangeb}) we also need to put a constraint on the unit price $p_i$ so that $b_i^\mathrm{min}<y^{-1}\left({p_ir_i}/{\left(\lambda r_i-p_i\right)}\right)$ always holds, which indicates 
\begin{equation}\label{pi_cons}
p_i<\frac{\lambda r_iy(b_i^\mathrm{min})}{y(b_i^\mathrm{min})+r_i}.
\end{equation}
Note that this constraint on $p_i$ is even stricter that the previous constraint $p_i<\lambda r_i$ noting that ${\lambda r_iy(b_i^\mathrm{min})}/{\left[y(b_i^\mathrm{min})+r_i\right]}<\lambda r_i$ holds.    

Until now we have derived the constraints on the leased spectrum as well as the unit price based on the conditions in (\ref{KKT1}) except the last two equalities. Now let's involve the last two equalities in (\ref{KKT1}), i.e., ${\mathrm{d} \mathscr{L}}/{\mathrm{d} {b_i}}=0$ and $\gamma_1\left[C_{th}-\overline{C}_{o}-C_{R,i}(b_i)\right]=0$. The equality ${\mathrm{d} \mathscr{L}}/{\mathrm{d} {b_i}}=0$ can be expressed as $\left(\lambda+\gamma_1\right)\mathcal{T}(b_i)=p_i$, 
where $\mathcal{T}(b_i)$ is the first derivative of $C_{R,i}(b_i)$ with $b_i$ given in (\ref{Tbinew}). When $C_{R,i}(b_i)= C_{th}-\overline{C}_{o}$ is satisfied, i.e., $b_i=b_i^\mathrm{min}$, ${\mathrm{d} \mathscr{L}}/{\mathrm{d} {b_i}}=0$ can be rewritten as $\left(\lambda+\gamma_1\right)\mathcal{T}(b_i^\mathrm{min})=p_i$.
This equation should hold under the condition that $\gamma_1\geq0$ and therefore $p_i\geq \lambda \mathcal{T}(b_i^\mathrm{min})$. Thus, invoking the previous constraint on $p_i$ given in (\ref{pi_cons}) we can get the spectrum demand $\mathscr{D}_i(p_i)=b_i^\mathrm{min}$ when $\lambda\mathcal{T}(b_i^\mathrm{min})\leq p_i<{\lambda r_iy(b_i^\mathrm{min})}/{[y(b_i^\mathrm{min})+r_i]}$. 

\sloppy On the other hand, when $C_{R,i}(b_i)> C_{th}-\overline{C}_{o}$ holds, i.e., $b_i>b_i^\mathrm{min}$, the equality $\gamma_1\left[C_{th}-\overline{C}_{o}-C_{R,i}(b_i)\right]=0$  necessitates $\gamma_1=0$. By substituting $\gamma_1=\gamma_2=0$ into $\left(\lambda+\gamma_1\right)\mathcal{T}(b_i)=p_i$, the equality ${\mathrm{d} \mathscr{L}}/{\mathrm{d} {b_i}}=0$ can be rewritten as (\ref{first_dri3}).
%\begin{equation}\label{first_dri3}
%{r_i\lambda\! \left\{\left[y(b_i)\right]^2\!\!+\!r_iy(b_i)\!-\!\frac{r_i}{\mathrm{ln}2}\left(1\!\!-\!2^{-y(b_i)}\right)\right\}}\!=\!p_i{\left[y(b_i)+r_i\right]^2}.
%\end{equation}
Hence the spectrum demand $\mathscr{D}_i(p_i)$ should be the root of the non-linear equation (\ref{first_dri3}) (if any) within the range of $b_i$ given in (\ref{rangeb}).
%, or equivalently, $\frac{p_ir_i}{\lambda r_i-p_i}<y(b_i) < y(b_i^\mathrm{min})$.
%Equation (\ref{first_dri3}) can be simplified as $\mathcal{T}(b_i)=p_i/\lambda$. 
Invoking the definition of $\mathcal{T}(b_i)$ given in (\ref{Tbinew}) and its first derivative given in (\ref{2der}), one can get that $\mathcal{T}(b_i)$ is a monotonically decreasing function with respect to $b_i$ with $\mathcal{T}(0)=r_i$ and $\mathcal{T}(+\infty)=0$. Therefore a single root of (\ref{first_dri3}), denoting as $b^\mathrm{root}_i$, always exists within the range given in (\ref{rangeb}) as long as the following inequality is satisfied 
\begin{equation}\label{Tinequ}
\mathcal{T}\left(y^{-1}\left(\frac{p_ir_i}{\lambda r_i-p_i}\right)\right)
<\frac{p_i}{\lambda}<\mathcal{T}(b_i^\mathrm{min}).
\end{equation} 
By substituting $b_i=y^{-1}\left(\frac{p_ir_i}{\lambda r_i-p_i}\right)$ into (\ref{Tbinew}), one can get 
\begin{equation}\label{Tlower}
\mathcal{T}\left(y^{-1}\left(\frac{p_ir_i}{\lambda r_i-p_i}\right)\right)=\frac{p_i}{\lambda}-\frac{\left(\lambda r_i-p_i\right)^2\left(1-2^{-\frac{p_ir_i}{\lambda r_i-p_i}}\right)}{\mathrm{ln}2\,\lambda^2r_i^2}. 
\end{equation}  
From the constraint of $p_i$ given in (\ref{pi_cons}) we know $p_i<\lambda r_i$ holds, thus the second term in the right hand side of (\ref{Tlower}) is positive, which indicates that the first inequality in (\ref{Tinequ}) always holds. Therefore, as long as the second inequality in (\ref{Tinequ}), i.e.,  ${p_i}<{\lambda}\mathcal{T}(b_i^\mathrm{min})$ is satisfied, a single positive root of (\ref{first_dri3}) $b^\mathrm{root}_i$ can be found within the range (\ref{rangeb}), which can be calculated numerically.    
%In Appendix we prove that a single positive root of (\ref{first_dri3}) in the given range of $b_i$ always exists as long as $p_i<\lambda \mathcal{T}(b_i^\mathrm{min})$ is satisfied. This spectrum demand can be expressed as $\mathscr{D}_i(p_i)=v_i\left(2^{y_\mathrm{root}}-1\right)^{-1}$ where $y_\mathrm{root}$ denotes the single positive root of (\ref{first_dri3}) with respect to $y(b_i)$. 

So far the optimization problem (\ref{opti1}) is solved completely as summarized in (\ref{demand}). {It is worth noting that when the unit price $p_i$ is above ${\lambda r_iy(b_i^\mathrm{min})}/{\left[y(b_i^\mathrm{min})+r_i\right]}$ or when $C_{th}-\overline{C}_{o}> v_i/\mathrm{ln}2$ holds, the KKT conditions in (\ref{KKT1}) cannot be all satisfied and the source will quit the spectrum trading resulting in zero spectrum demand $\mathscr{D}_i(p_i)=0$.} 
\end{proof}
It is interesting to investigate the behaviour of the spectrum demand $\mathscr{D}_i(p_i)$ with respect to the increase of the unit price $p_i$. Since $\mathcal{T}(b_i)$ is a monotonically decreasing function, larger $p_i$ will result in smaller root of equation (\ref{first_dri3}), i.e., $b^\mathrm{root}_i$. Therefore, the demand function given in (\ref{demand}) is a monotonically decreasing function of the unit price in low $p_i$ regime. This is a reasonable result considering that with the increase of unit spectrum price, the source will require less spectrum demand due to the increase of the cost. However, with the further increase of $p_i$, the demand saturates at a fixed value $b_\mathrm{min}$ which is the minimal bandwidth that the source wants to borrow in order to achieve the data rate requirement. In this stage, the source will only request to borrow this minimum required bandwidth, since the unit price is too high so that borrowing a bit more bandwidth will result in the reduction of its utility. When the price keeps increasing so that even with the minimal leased bandwidth the achievable utility is negative, the source will quit the game and demand spectrum will drop to zero. 

%Based on the above discussion, in order to establish this demand function, two non-linear equations have to be solved, using methods such as bisection method. The first is (\ref{bmin}) and the second is (\ref{A=B2}). The solution of the former can determine $b_i^\mathrm{min}$ the minimal shared bandwidth that the source should borrow from the $i$th relay to achieve the data rate requirement and the solution of the latter gives $y_\mathrm{root}$ for a given price $p_i$. For each channel state, the first non-linear equation just needs to be solved once, however, the latter needs to be solved for many times with different price samples. To simplify the complexity, an approximated $y_\mathrm{root}$ can be derived analytically. When SNR of the $\mathcal{S}-\mathcal{R}_i$ link is much larger than $1$, i.e., $v_i/b_i\gg1$, the function $\mathcal{B}$ can be approximated as a fixed value $r_i/\mathrm{ln}2$ and the root of Eq. (\ref{A=B2}) can be expressed as
%\begin{equation}
%y_\mathrm{root}^a=\frac{-\left(r_i-\frac{2p_i}{\lambda}\right)+\sqrt{r_i^2+\frac{4}{\mathrm{ln}2}\left(r_i-\frac{p_i}{\lambda}\right)}}{2\left(1-\frac{p_i}{\lambda r_i}\right)}.
%\end{equation}
%Note that generally the demand function achieved using $y_\mathrm{root}^a$ above is larger the optimal demand, however, it becomes close to the optimal demand function with the increase of the SNR of the link $\mathcal{S}-\mathcal{R}_i$. 

\subsection{Utility of the Relays and the Supply Function}\label{suppfunc}
As mentioned in Section \ref{System Model}, the RF nodes considered in this work have their own data to transmit to the destination. Although they can gain extra revenue by lending spectrum to the source, they may experience the QoS degradation of their connected UEs. Assuming that a RF node serves $M_i$ UEs and each with a constant data rate requirement of $R_i^\mathrm{ur}$, the utility of each RF node can be expressed as \cite{Niyato08}
\begin{equation}\label{uti_relay}
\mathcal{P}_i=b_ip_i+c_1M_i-c_2M_i\left(R_i^\mathrm{ur}-\frac{\left(W_i-b_i\right)r_i}{M_i}\right)^2,
\end{equation}
where $c_1$ denotes the constant weight for the revenue of serving each local connection whereas $c_2$ denotes the constant weight for the cost of QoS degradation, the first term refers to the spectrum lending revenue based on linear pricing, the second term denotes the income of providing the service of local data transmission, and the third term is the cost due to the QoS degradation. In (\ref{uti_relay}), it is assumed that the RF node charges a fixed fee for serving every connected UE to communicate with MBS so that its income of offering local service can be expressed as $c_1M_i$. In addition, with the assumption that the remaining data rate is uniformly allocated to $M_i$ UEs, the available data rate for each UE is ${\left(W-b_i\right)r_i}/{M_i}$. Therefore, the cost induced by the QoS degradation of each UE can then be expressed as $c_2\left(R_i^\mathrm{ur}-\left(W_i-b_i\right)r_i/M_i\right)^2$. This cost can be treated as the discount offered to the UE because of the spectrum lending to the source. The quadratic form of the QoS degradation cost has been widely used in literature \cite{Lin11,Niyato08,Niyato083}, which indicates that the dissatisfaction of the UEs increases quadratically with the gap between the required data rate and the actual data rate. 
%It is worth emphasizing that the utility function given in (\ref{uti_relay}) is a general form which can accurately model the application of the system in the wireless backhauling plotted in Fig. \ref{system}. In this application, each SBS severs several UEs in the small cell with $M_i$ denoting the number of ongoing UEs. However, for some other applications where the RF nodes are used for point-to-point high-speed data transmission to the destination, each RF node actually severs only one local connection and hence one should substitute $M_i=1$ into (\ref{uti_relay}). 

{To establish a RF relay link $\mathcal{S}-\mathcal{R}_i-\mathcal{D}$,  in our proposed spectrum trading game each RF has to solve an optimization problem to get the optimal size of the leased spectrum for any given price $p_i$, which is the so-called spectrum supply function.} This optimization problem is given by
\begin{align}\label{opti2}
\underset{b_i}{\mathrm{max}}& \quad\mathcal{P}_i=b_ip_i+c_1M_i-c_2M_i\left(R_i^\mathrm{ur}-\frac{\left(W_i-b_i\right)r_i}{M_i}\right)^2,\\\nonumber
\mathrm{s.t.}& \quad b_i>0, \quad b_i\leq W_i, \quad \mathcal{P}_i>\mathcal{P}_i^\mathrm{no},
\end{align}
where $\mathcal{P}_i^\mathrm{no}$ denotes the RF node's gained utility when the RF node does not lend its spectrum to the source given by 
\begin{equation}
\mathcal{P}_i^\mathrm{no}=c_1M_i-c_2M_i\left[\left(R_i^\mathrm{ur}-\frac{W_ir_i}{M_i}\right)^+\right]^2.
\end{equation} 
Note that $x^+=\mathrm{max}(x,0)$. The first condition in (\ref{opti2}) indicates that the leased spectrum is positive, the second inequality condition means that the leased spectrum should not exceed the total licensed spectrum of the RF node, i.e., $W_i$, and the third inequality represents that by lending the spectrum the RF node is able to enhance its utility, otherwise the RF node will quit the game due to the loss of utility. The solution of the optimization problem (\ref{opti2}) gives the optimal spectrum supply denoted as $\mathscr{S}_i(p_i)$  as a function of the price $p_i$. 

{\begin{Proposition}
The solution to the optimization problem (\ref{opti2}) is given by
\begin{equation}\label{supply}
\mathscr{S}_i(p_i)=
\left\{
\begin{array}{ll}
W_i-\frac{M_iR_i^\mathrm{ur}}{r_i}+\frac{M_ip_i}{2c_2r_i^2} ,\, &p_{L}<p_i<2c_2r_iR_i^\mathrm{ur},\\
W_i, \quad &p_i\geq2c_2r_iR_i^\mathrm{ur},\\
0,    \quad & \mathrm{otherwise},
\end{array}
\right.
\end{equation}
where 
\begin{equation}\label{pL}
p_{L}=\left(2c_2r_i\left[R_i^\mathrm{ur}-\frac{r_iW_i}{M_i}\right]\right)^+.
\end{equation}
\end{Proposition}}  

\begin{proof}
To solve the optimization problem given in (\ref{opti2}), we should firstly check the convexity of the objective function. By taking the second derivative of the objective function with respect to $b_i$, one can get ${\mathrm{d}^2 \mathcal{P}_i}/{\mathrm{d} b_i^2}=-2c_2r_i^2/M_i$ which means that the objective function $\mathcal{P}_i$ is concave. Therefore, we can again use KKT conditions to get the optimal spectrum supply $b_i$. The Lagrangian associate with problem is 
\begin{equation}
\mathscr{L}=-\mathcal{P}_i+\gamma_1\left(\mathcal{P}_i^\mathrm{no}-\mathcal{P}_i\right)-\gamma_2b_i-\gamma_3\left(W_i-b_i\right).
\end{equation}
The corresponding KKT conditions for this optimization problem can be expressed as 
\begin{align}\label{KKT2}
&b_i>0, \quad b_i\leq W_i,\quad \mathcal{P}_i>\mathcal{P}_i^\mathrm{no}, \quad \gamma_1,\gamma_2,\gamma_3\geq 0, \\\nonumber &\gamma_1\left(\mathcal{P}_i^\mathrm{no}-\mathcal{P}_i\right)=0, \, \frac{\mathrm{d} \mathscr{L}}{\mathrm{d} {b_i}}=0, \, \gamma_2b_i=0, \,\gamma_3\left(W_i-b_i\right)\!=0.
\end{align}
Since the inequality $b_i>0$ should hold, the equality $\gamma_2b_i=0$ necessitates $\gamma_2=0$. Similarly, $\mathcal{P}_i>\mathcal{P}_i^\mathrm{no}$ and $\gamma_1\left(\mathcal{P}_i^\mathrm{no}-\mathcal{P}_i\right)=0$ necessitate $\gamma_1=0$. After some algebraic manipulations, the equality ${\mathrm{d} \mathscr{L}}/{\mathrm{d} {b_i}}=0$ can be expressed as
\begin{equation}\label{first_deri2}
\frac{\mathrm{d} \mathscr{L}}{\mathrm{d} {b_i}}=- \left[p_i-2c_2r_i\left(R_i^\mathrm{ur}-\frac{\left(W_i-b_i\right)r_i}{M_i}\right)\right]+\gamma_3=0.
\end{equation}

Assuming $b_i<W_i$, in order to make sure the equality $\gamma_3\left(W_i-b_i\right)=0$ in KKT conditions (\ref{KKT2}) is satisfied, the parameter $\gamma_3$ should be zero. Substituting $\gamma_3=0$ into (\ref{first_deri2}), one can get the optimal spectrum supply function denoted as $\mathscr{S}_i(p_i)$ given by
\begin{equation}\label{supply_new}
\mathscr{S}_i(p_i)=W_i-\frac{M_iR_i^\mathrm{ur}}{r_i}+\frac{M_ip_i}{2c_2r_i^2}.
\end{equation} 
In order to ensure that the rest KKT conditions are all satisfied, this spectrum supply should also satisfy $0<\mathscr{S}_i(p_i)<W_i$ and $\mathcal{P}_i>\mathcal{P}_i^\mathrm{no}$. By substituting (\ref{supply_new}) into these two inequalities and after some manipulations, we can get that a constraint on the given unit price $p_i$ should be satisfied as $p_{L}<p_i<2c_2r_iR_i^\mathrm{ur}$ where $p_{L}$ is given in (\ref{pL}). However, outside this range of given price the KKT conditions cannot be all guaranteed and hence no optimal spectrum supply exists when $b_i<W_i$. 

Secondly, let's consider the case when the optimal spectrum supply equals to the total bandwidth, i.e., $b_i=W_i$. Substituting this into (\ref{first_deri2}), one can get  $-\left(p_i-2c_2r_iR_i^\mathrm{ur}\right)+\gamma_3=0$. Considering that $\gamma_3$ is non-negative, this equality results in a constraint for the price as $p_i\geq 2c_2r_iR_i^\mathrm{ur}$. It can be easily shown that when this constraint on $p_i$ is satisfied, the inequality $\mathcal{P}_i>\mathcal{P}_i^\mathrm{no}$ also holds. Therefore, as long as $p_i\geq 2c_2r_iR_i^\mathrm{ur}$ is met, we have the optimal spectrum supply $\mathscr{S}_i(p_i)=W_i$.           
%To further ensure that the inequality $\mathcal{P}_i>\mathcal{P}_i^\mathrm{no}$ holds, one can substitute $b_i=W_i$ into (\ref{uti_relay}) and get that when $M_iR_i^\mathrm{ur}\geq Wr_i$, $p_i\geq2c_2r_iR_i^\mathrm{ur}-c_2W_ir_i^2/M_i$ should hold, whereas when $M_iR_i^\mathrm{ur}< W_ir_i$, $p_i>{c_2M_i\left(R_i^\mathrm{ur}\right)^2}/{W_i}$ should hold instead. However, both constraints on $p_i$ is less strict than $p_i\geq 2c_2r_iR_i^\mathrm{ur}$, therefore, as long as $p_i\geq 2c_2r_iR_i^\mathrm{ur}$ is satisfied, we have the optimal spectrum supply $\mathscr{S}_i(p_i)=W_i$.        
{Note that when the unit price $p_i$ is less than $p_L$, the KKT conditions in (\ref{KKT2}) cannot be all satisfied and the relay will quit the spectrum trading resulting in zero spectrum supply, i.e., $\mathscr{S}_i(p_i)=0$.} 
\end{proof}
The optimal spectrum supply given in (\ref{supply}) illustrates that with the increase of the unit price, the spectrum supply will firstly be zero and then increase with the increase of the unit price. Finally when the price is high enough, the relay is pleased to lend all of its licensed spectrum $W_i$ to gain higher revenue. 

%Note that in practical RF nodes, the condition $M_iR_i^\mathrm{ur}<2r_iW_i$ is usually satisfied with high probability.          

\subsection{{Spectrum trading Process and Relay Selection}}\label{relay selection}
We have derived the demand function of the source given in (\ref{demand}) and the supply function of the relays given in (\ref{supply}). The remaining problem is how to reach the market-equilibrium price using these functions. The market-equilibrium is defined as the situation where the supply of an item is exactly equal to its demand so that there is neither surplus or shortage in the market and the price is hence stable in this situation \cite{Niyato08,Niyato082}. As discussed in Section \ref{demandfunc}, the spectrum demand function is a decreasing function with respect to the price in low unit price regime, which means higher price will result in less spectrum demand due to high cost. One the other hand, as discussed in Section \ref{suppfunc}, supply function is an increasing function with respect to the price, which in turn implies that higher price will lead to more spectrum supply due to the higher revenue. Therefore, the cross point of these two functions (if exists), i.e., 
\begin{equation}\label{marketeq}
\mathscr{D}_i(p_i ^*)=\mathscr{S}_i(p_i^*),
\end{equation}
gives the market-equilibrium price $p^*_i$ where the spectrum demand and supply are balanced. In this equilibrium, both source and RF node are happy with the price and size  of the  leased spectrum. 
%The relay might seek to increase the price to get higher revenue, however, this action will lead to less spectrum demand from the source and in turn reduces the relay's profit. However, 
Since an analytical expression of the root for (\ref{marketeq}) cannot be derived, we solve this equation numerically using bisection method. It is possible that the root of (\ref{marketeq}) does not exist which means no market-equilibrium price can be reached and thus the spectrum trading between the source and the RF node cannot be established.          

So far we mainly focused on the spectrum trading process between the source $\mathcal{S}$ and the $i$th RF relay $\mathcal{R}_i$. However, as illustrated in Fig. \ref{system}, there may be multiple surrounding RF node candidates that can act as the relay for the source. Therefore, a relay selection method needs to be developed. When $\overline{C}_{o}<C_{th}$ is satisfied due to the presence of adverse weather conditions, the source will notify the distributed $N$ available RF nodes and broadcast some information including its minimal required data rate and the current average FSO link capacity. Assuming that all RF nodes know the form of the source demand function given in (\ref{demand}), each RF node is able to generate the exact demand function of the source. In addition, based on its own traffic load information such as the number of connected UEs and their data rate requirements, every RF node can also generate its own supply function using the equation given in (\ref{supply}). With both calculated demand and supply functions, every RF node could calculate its proposed market-equilibrium unit spectrum price $p^*_i$ using (\ref{marketeq}) as well as the corresponding optimal portion of leased spectrum to offer. The RF nodes will then send this information back to the source. Having the proposed price and the optimal leased spectrum from each RF node, the source will be able to calculate its own utility using (\ref{uti_sour}) to finally select the RF node which provides it with the maximal utility. After notifying the selected RF node, the source is able to use the leased bandwidth to establish RF dual-hop relay link. Based on the design of the spectrum trading game, this RF relay link together with the FSO link enables the source to transmit data to the destination with a rate above the data rate requirement $C_{th}$. Note that the proposed spectrum trading game and relay selection will be restarted after each time interval $T_u$.

\section{Numerical Result Analysis} \label{Numerical Result}
In this section, we present some simulation results for our proposed system in the application of wireless backhauling as plotted in Fig. \ref{system}. Unless otherwise stated, the values of the system parameters used for the numerical simulations are listed in the Table \ref{table} \cite{Jamali16,zhang09,he09}. Note that for simplicity it is assumed that each RF relay has the same amount of total licensed bandwidth $W_i=W,\, \forall i\in\{1,\ldots,N\}$ and the RF fading coefficient $h_{R,i}^{(t)}$ is modelled as Rayleigh distribution \cite{Dahrouj15,Siddique15}. In addition, the transmitted RF power at the source and the relays are considered to be equal, i.e., $P_{R}^\mathcal{S}=P_{R}^\mathcal{R}=P_{R}$.  The property of the market-equilibrium pricing in the absence of RF channel fading will be firstly presented and the communication performance improvement of employing the proposed spectrum trading strategy will then be discussed.  
\begin{table}
	\large
	\renewcommand{\arraystretch}{1.5}
	\caption{The Parameter Setting  \cite{Jamali16,zhang09,he09}}
	\label{table}
	\centering
	\resizebox{\textwidth}{!}{\begin{tabular}{ll}
{\begin{tabular}{|c|c|c|}
		\hline 
		\multicolumn{3}{|c|}{FSO Link}  \\
		\hline \hline
		Symbol & Definition & Value\\
		\hline\hline
		$d$ & Receiver aperture diameter & $5$ cm \\
		\hline
		$\rho$ & Responsivity of FSO photodetector & $0.5$ $\mathrm{V}^{-1}$\\
		\hline
		$L_{SD}$ & Distance between the source and destination & $1000$ m\\
		\hline
		$\phi$ & Beam divergence angle & $3.5$ mrad \\
		\hline
		$C_n^2$ & Refraction structure index & $5\times10^{-14}$ $\mathrm{m}^{-2/3}$ \\
		\hline
		$\lambda_{o}$ & Laser wavelength & $1550$ nm\\
		\hline  
		$\sigma_{o}^2$ & Noise variance at FSO receiver & $10^{-14}$ $\mathrm{A}^2$\\
		\hline
		$P_{o}$ & Optical transmission power at the source & $20$ mW\\ 
		\hline
		$\lambda$ & Utility gain per unit data rate & $1$ $\mathrm{Mbps}^{-1}$ \\
		\hline
		$W_{o}$ & Bandwidth of FSO link & $1$ GHz \\
		\hline
		$C_{th}$ & Minimal data rate requirement & $80$ Mbps\\ 
		\hline
	\end{tabular}}
&
{\begin{tabular}{|c|c|c|}
	    \hline 
		\multicolumn{3}{|c|}{RF Link}  \\
		\hline \hline
		Symbol & Definition & Value\\
		\hline\hline
		$\lambda_{R}$ & RF wavelength & $85.7$ mm\\
		\hline
		$G_{TX}$, $G_{RX}$ & Antenna Gain & ($10$,$10$) dBi\\
		\hline
		$L_\mathrm{ref}$ & Reference distance of the RF link & $80$ m\\
		\hline
		$L_{R,i}^{(1)}$ & Distance between source and RF nodes & $600$ m\\
		\hline
		$L_{R,i}^{(2)}$ & Distance between RF nodes and destination & $600$ m\\
		\hline
		$P_{R}^\mathcal{S}$, $P_{R}^\mathcal{R}$ & RF transmitter power at the source and the RF nodes & $0.2$ W\\
		\hline
		$N_0$ & Noise power spectral efficiency at RF receiver & $-114$ dBm/MHz\\
		\hline
		$W$ & Licensed spectrum for the $\mathcal{R}_i-\mathcal{D}$ link  & $20$ MHz\\
		\hline
		$R_i^\mathrm{ur}$ & Data rate requirement per user equipment & $3$ Mbps\\
		\hline 
		$\delta$ & RF path-loss exponent & $3.5$ \\
		\hline
		$c_1,c_2$ & Constant weights for the utility of the RF nodes & $(1,0.5)$\\
		\hline
	\end{tabular}}
	\end{tabular}}
			\squeezeup
			\squeezeup
	\end{table}

\subsection{Market-Equilibrium Pricing}
\begin{figure}[!t]
		\squeezeup
		\squeezeup
	\centering
	\includegraphics[width=0.85\textwidth]{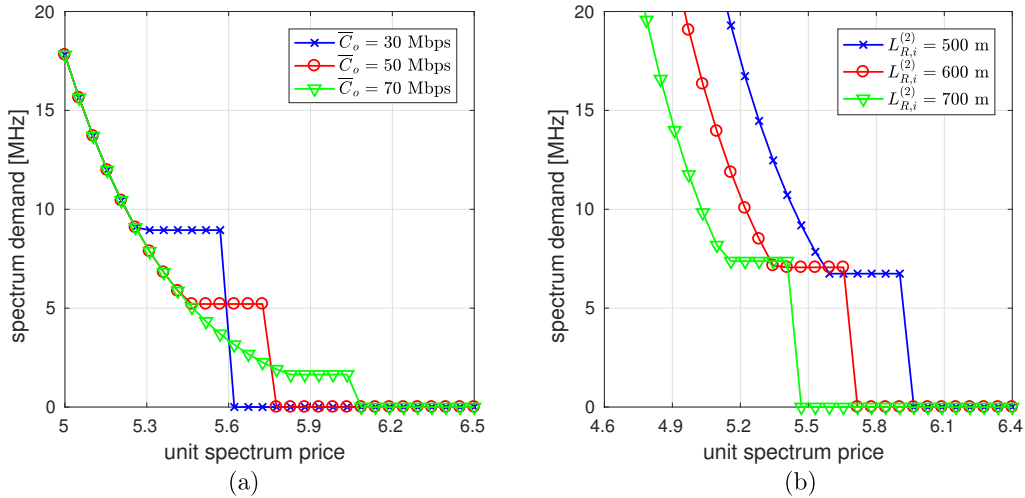}
	\caption{The spectrum demand versus the unit spectrum price: (a) for different FSO link conditions; (b) for different distance between the RF node and destination with average FSO link capacity $\overline{C}_{o}=40$ Mbps.  }\label{F1_F2_combin}
	\squeezeup
	\squeezeup
\end{figure}

Figure \ref{F1_F2_combin}(a) plots the source demand function (\ref{demand}) with respect to the unit spectrum price $p_i$ with various average FSO link capacities ($\overline{C}_{o}$). This figure shows that with the increase of the unit price $p_i$, the bandwidth demand firstly decreases and saturates at a fixed value which denotes the minimum required bandwidth. In addition, with the increase of $\overline{C}_{o}$ less bandwidth is required to achieve $C_{th}$ which results in reduction of the saturation level. For instance, when $\overline{C}_{o}=30$ Mbps, the minimum required bandwidth is $8.94$ MHz, however, when $\overline{C}_{o}=70$ Mbps, the corresponding minimum required bandwidth is only $1.64$ MHz. With further increase of the unit spectrum price, since the source cannot achieve a positive utility by even buying the minimal bandwidth, it will quit the game and the bandwidth demand will reduce to zero. One can observe that when the FSO link is in a better condition, the source has a higher tolerance to the increase of the unit spectrum price. For example, the spectrum demand drops to zero on $p_i=5.57$ when $\overline{C}_{o}=30$ Mbps, but the corresponding price increases to $6.03$ when $\overline{C}_{o}=70$ Mbps. This is because better FSO links require less leased bandwidth and the cost of borrowing this amount of bandwidth reduces accordingly and hence the source is able to accept higher unit spectrum price.
  
Besides the condition of FSO channel, the behaviour of the demand function is also associated with the locations of the RF nodes, which determines the data rate of the RF relay link. Figure \ref{F1_F2_combin}(b) shows the effect of RF node locations to the demand function. The distance between the source and RF node is fixed at $L_{R,i}^{(1)}=600$ m, whereas the the distance between the RF node and destination $L_{R,i}^{(2)}$ varies. The same results can also be observed when $L_{R,i}^{(1)}$ changes. This figure shows that, for a given low unit spectrum price, lower $L_{R,i}^{(2)}$ results in higher spectrum demand. For example, the spectrum demand when $L_{R,i}^{(2)}=600$ m is $18$ MHz with a unit price $5$, however, the corresponding bandwidth demand when $L_{R,i}^{(2)}=700$ m is only $11$ MHz. This is because smaller distance between the RF node and destination results in better channel condition in $\mathcal{R}_i-\mathcal{D}$ link.  The source is then happy to buy more bandwidth, which can result in higher data rate to the destination and hence higher utility. 
%In addition, one can also get that with less distance between the RF node and destination, the minimal required bandwidth to achieve data rate threshold decreases and the source also becomes more tolerable to the increase of unit price. 
%For instance, when $L_{R,i}^{(2)}=700$ m, the minimal required bandwidth is $7.3$ MHz and the maximal tolerant unit spectrum price is $5.4$, however, when $L_{R,i}^{(2)}$ decreases to $500$ m, the corresponding bandwidth and price is $6.7$ MHz and $5.9$, respectively.     

%\begin{figure}[!t]
%	\centering
%	\includegraphics[width=0.55\textwidth]{F2_market_demand_location.eps}
%	\caption{The spectrum demand versus the unit price for various distance between the RF node and destination under average FSO link capacity $\overline{C}_{o}=40$ Mbps. }\label{market_demand_fading}
%\end{figure}

\begin{figure}[!t]
		\squeezeup
		\squeezeup
	\centering
	\includegraphics[width=0.85\textwidth]{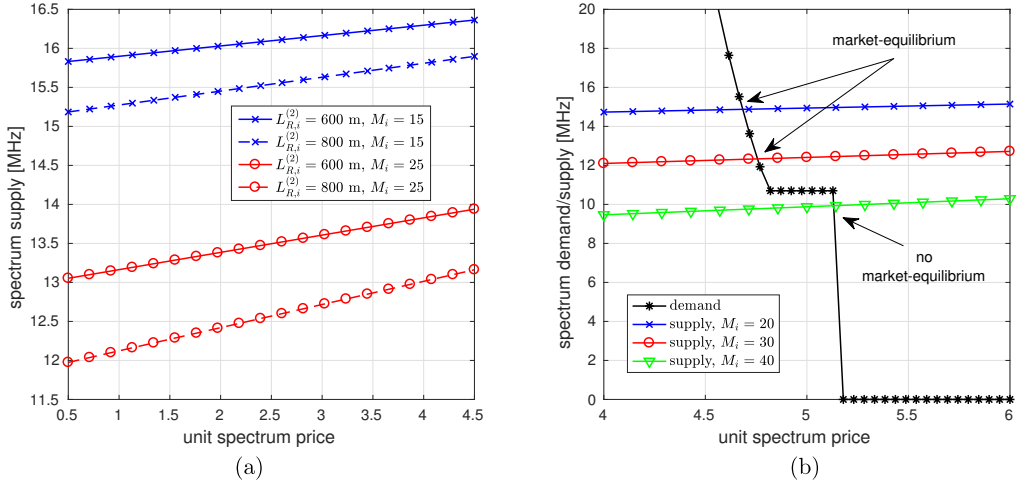}
	\caption{(a) The spectrum supply versus the unit spectrum price for different numbers of connected UEs in the small cell and link distance between the relay and destination; (b) The demand and supply functions with respect to the unit spectrum price for different number of connected UEs in the small cell where $C_{o}=25$ Mbps and $L_{R,i}^{(1)}=L_{R,i}^{(2)}=700$ m.}\label{F3_F4_combin}
		\squeezeup
		\squeezeup
\end{figure}

From the supply function given in (\ref{supply}), we know that the behaviour of the supply function is associated with both the location of the RF node and the number of connected UEs in RF node. Figure \ref{F3_F4_combin}(a) plots the supply function versus the unit price with various number of connected UEs and distance between the RF node and destination. It is presented that with the increase of the unit price, the bandwidth supply increases. For instance, when $L_{R,i}^{(2)}=600$ m and $M_i=25$, the bandwidth supply is $13.1$ MHz with a given spectrum price $1$, whereas the corresponding bandwidth increases to $13.7$ MHz when the spectrum price increases to $3.5$. 
%Note that the increase of the spectrum supply from zero and its saturation at $W=20$ MHz are out of the investigated range of Fig. \ref{market_sup_new}. 
In addition, it is also obvious that with the increase of the distance $L_{R,i}^{(2)}$ and the number of connected local UEs $M_i$, the spectrum supply for a given price decreases due to the less total data rate to the destination and higher traffic load, respectively. 

%For instance, the RF node prefers to offer a portion of bandwidth $16.1$ MHz for a unit price $2.5$ with $L_{R,i}^{(2)}=600$ m and $M_i=15$. However, when $L_{R,i}^{(2)}$ increases to $800$ m and $M_i$ increases to $25$, the corresponding spectrum supply decreases to $15.5$ MHz and $13.5$ MHz, respectively.                
       
Figure \ref{F3_F4_combin}(b) shows both supply and demand functions with respect to the unit spectrum price under different number of connected UEs in the small cell. The market-equilibrium price is given by the cross-point when the supply function meets the demand function. One can observe that the market-equilibrium only exists for a range of connected UE numbers. For example, market-equilibrium prices for $M_i=20$ and $M_i=30$ are $p^*_i=4.68$ and $p^*_i=4.75$, respectively. However, no market-equilibrium price exists for $M_i=40$ due to the high traffic load for RF node. We would like to emphasize that beside the number of connected UEs, the existence of market-equilibrium is also associated with many other conditions, i.e., the FSO link condition, the distance between the source (RF node) and RF node (destination), and also the fading conditions of RF links.

\subsection{Performance Improvement}
Now let's consider the performance improvement achieved by the proposed system in the wireless backhauling application considering Rayleigh fading in RF links. We assume that the data rate requirement of each UE in the small cell $R_i^\mathrm{ur}$ is a constant, however, the traffic load of the SBS varies due to the random number of connected UEs $M_i$, which is modelled as a Poisson distributed random variable with mean $\upsilon_M$.   

\begin{figure}[!t]
		\squeezeup
		\squeezeup
	\centering
	\includegraphics[width=0.86\textwidth]{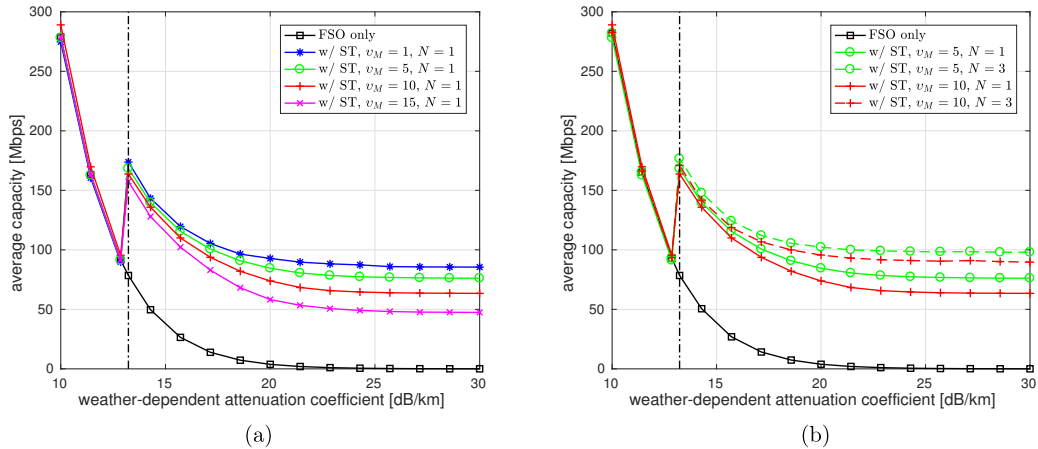}
	\caption{The average capacity versus the weather-dependent attenuation coefficient of the FSO link $\kappa$: (a) with a single RF node involved in the spectrum trading, i.e., $N=1$, and various average number of connected UEs $\upsilon_M$ in the small cell; (b) with various $\upsilon_M$ under different number of surrounding small cells $N$. The vertical dashed-dotted line represents the $\kappa$ which results in an average capacity of FSO link equal to the threshold $C_{th}$. ST: spectrum trading. }\label{F5_F6_combin}
		\squeezeup
		\squeezeup
\end{figure}
Figure \ref{F5_F6_combin}(a) presents the average capacity versus weather-dependent attenuation coefficient $\kappa$ with various $\upsilon_M$ with and without the spectrum trading. Note that in this figure a single RF node locating at $L_{RF,i}^{(1)}=L_{RF,i}^{(2)}=600$ m is considered, i.e., $N=1$, and in order to get accurate average capacity performance, $3\times 10^3$ samples of channel realizations are generated for each $\kappa$. From Fig. \ref{F5_F6_combin}(a) one can observe that with the increase of $\kappa$ the average capacity for FSO-only link decreases significantly and when $\kappa$ is above $20$ dB/km the capacity of FSO link becomes negligible. However, in the presence of proposed spectrum trading, as long as the average FSO capacity is less than the data rate requirement $C_{th}=80$ Mbps (or equivalently when $\kappa>13.23$ dB/km), the average capacity of the system can be significantly improved by means of establishing the dual-hop RF/FSO hybrid link. For instance, an average capacity of $130$ Mbps can be achieved when $\kappa=15$ dB/km and $\upsilon_M=1$ when spectrum trading is employed, however, for FSO system without spectrum trading the corresponding average capacity is only $40$ Mbps. In addition, it is also presented that the performance of systems with smaller $\upsilon_M$ outperforms those with larger $\upsilon_M$ as expected, because of the lower probability of high traffic loads. Furthermore, with the increase of attenuation coefficient $\kappa$, the average capacities of systems with spectrum trading decrease and finally saturate at fixed values. For example, the asymptotic average capacities at high $\kappa$ are $76.15$ and $63.42$ for systems with $\upsilon_M=5$ and $\upsilon_M=10$, respectively. This is because when $\kappa\to \infty$, the FSO link is totally non-functional and the throughput from the source to the destination thoroughly relies on the dual-hop RF relay link, which provide the source with a fixed average capacity.               
%\begin{figure}[!t]
%	\centering
%	\includegraphics[width=0.5\textwidth]{F6_average_capa_multi_fading_new.eps}
%	\caption{The average capacity versus the weather-dependent attenuation coefficient of the FSO link $\kappa$ with various $\upsilon_M$ under different number of surrounding small cells $N$. The vertical dashed-dotted line represents the $\kappa$ which results in an average capacity of FSO link equal to the threshold $C_{th}$. ST: spectrum trading.}\label{capa_mean_poiss_double}
%\end{figure}
\begin{figure}[!t]
		\squeezeup
		\squeezeup
	\centering
	\includegraphics[width=0.92\textwidth]{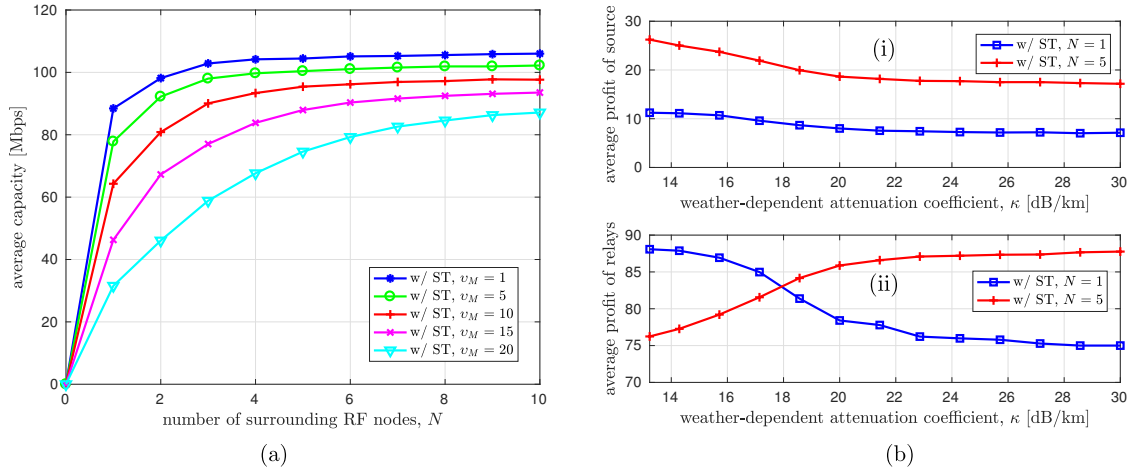}
	\caption{(a) The asymptotic average capacity versus the number of surrounding SBS $N$ in high $\kappa$ regime for various $\upsilon_M$. (b) The gained profit versus the weather-dependent attenuation coefficient of the FSO link with $\upsilon_M=5$ for various $N$: (i) the profit of the source; (ii) the total profit of relays. ST: spectrum trading.}\label{F9_F10_combin}
		\squeezeup
		\squeezeup
\end{figure}

Figure \ref{F5_F6_combin}(b) plots the performance of the average capacity with various number of surrounding RF small cells $N$. It is shown that with the increase of $N$, better average capacity performance can be achieved. For instance, by increasing $N=1$ to $N=3$, the average capacity increases from $84.57$ Mbps to $102.30$ Mbps for $\upsilon_M=5$ and $\kappa=20$ dB/km. Accordingly, when $\upsilon_M=10$ the average capacity rises from $73.90$ Mbps to $95.62$ Mbps. This is because when the FSO link is unavailable due to the adverse weather conditions, with larger $N$ the source has more SBS candidates to choose and when one SBS is in high traffic load and market-equilibrium spectrum trading cannot be established, the source is able to look for the other SBSs to buy the RF spectrum. Therefore the probability of establishing dual-hop RF relay link increases with the increase of $N$, which leads to better performance. In fact, the effect of relay selection in this work is similar to the selection combining in the context of spatial diversity where the most favourable signal at the receiver among several received signals is selected for decoding. 

To further illustrate the effects of involving more RF nodes into the spectrum trading game, the asymptotic average capacity in high $\kappa$ versus $N$ is plotted in Fig. \ref{F9_F10_combin}(a). It can be observed that for all systems with various traffic load conditions, a significant capacity improvement can be achieved even when a single RF node is employed, especially for those with less $\upsilon_M$. However, the improvement of the average capacity with the increase of $N$ is more significant for systems with higher traffic loads in RF nodes, i.e., larger $\upsilon_M$. For example, with $\upsilon_M=1$, compared to the system with a single surrounding small cell ($N=1$), an average capacity improvement around $17.50$ Mbps can be achieved when $10$ small cells are involved. However, for the system with higher traffic load in small cells, i.e., $\upsilon_M=20$, the capacity improvement of increasing $N$ from $1$ to $10$ is $55.57$ Mbps.               

%\begin{figure}[!t]
%	\centering
%	\includegraphics[width=0.6\textwidth]{F10_profit.eps}
%	\caption{The gained profit versus the weather-dependent attenuation coefficient of the FSO link with $\upsilon_M=5$ for various $N$: (a) the profit of the source; (b) the total profit of relays. ST: spectrum trading.   }\label{F10_profit}
%\end{figure}

Figure \ref{F9_F10_combin}(b) plots the profit of the source and relays versus the weather-dependent attenuation coefficient of the FSO link. It is presented that with the increase of $\kappa$, the profit of the source decreases. This is because with higher $\kappa$, the market-equilibrium is harder to achieve, which results in lower probability of establishing the hybrid dual-hop RF/FSO link and hence smaller average profit of source. However, the source always benefits from employing more surround RF nodes due to the increase of the probability of hybrid link establishment by means of diversity. Therefore, a significant profit improvement of the source can be observed in the upper figure of Fig. \ref{F9_F10_combin}(b) by increasing $N=1$ to $N=5$. On the other hand, the behaviour of the profit of relays with $\kappa$ varies with $N$ as shown in the lower figure of Fig. \ref{F9_F10_combin}(b). Note that here we consider the total profit of the relays when several RF nodes are involved. When $N$ is small (e.g., $N=1$), the profit of relays decreases as that of the source with the increase of $\kappa$, which is again due to the decrease of the hybrid link establishment probability. However, with a relatively large $N$ (e.g., $N=5$), this probability is improved which does not limit the profit of the relays any more. In this scenario, the relays can get higher profit with the increase of $\kappa$, because the source has to borrow more bandwidth  due to the non-functional FSO link. In addition, in lower $\kappa$ regime, the market-equilibrium is relatively easy to be achieved and with higher $N$ the source has more candidates to choose and is more likely to select the one with lower spectrum price, which leads to lower profit of relays. As a result, when $\kappa$ is small, the relays' profit when $N=1$ is even higher than that of $N=5$. It is worth mentioning that based on our simulation the profit of the source also always benefits from less traffic loads (lower $\upsilon_M$) as expected.

\begin{figure}[!t]
	\centering
	%	\hspace*{-0.6cm}  
	\includegraphics[width=0.65\textwidth]{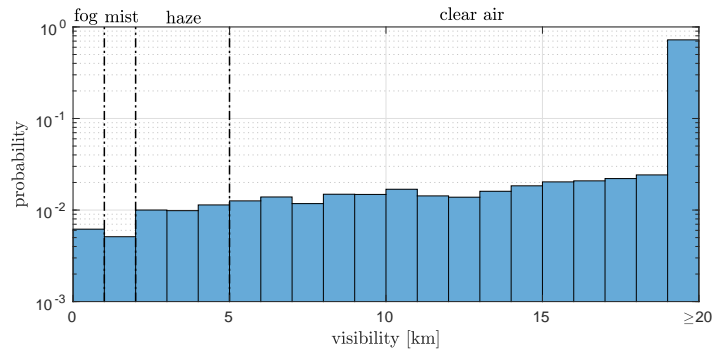}
	\caption{The histogram of the hourly visibility for Edinburgh from January $2016$ to June $2017$. }\label{F8_histo}
		\squeezeup
		\squeezeup
\end{figure}
Finally, let's consider the application of our proposed system in a realistic channel model based on the climate data for the city of Edinburgh. The measured hourly visibility data of the Edinburgh Gogarbank weather station for January $2016$ to June $2017$ (totally $\mathscr{H}_\mathrm{tot}=13128$ hours) is provided by the United Kingdom's national weather service the Meteorological Office. Fig. \ref{F8_histo} shows the histogram of the hourly visibility. One can see that the probability of fog events (with visibility less than $1$ km) which might severely degrade the FSO link performance is quite low (around $6\times 10^{-3}$). This reveals that the hybrid RF/FSO links proposed in previous works \cite{Abdul10,zhang09} with RF links to be active continuously are not necessarily the most economical choice, since in most of the time FSO links perform well enough. On the other hand, the proposed system in this work benefits from establishing the hybrid link only when the low-frequent adverse weather conditions appear.                 

\begin{figure}[!t]
	\centering
	%	\hspace*{-0.6cm} 
	\includegraphics[width=0.65\textwidth]{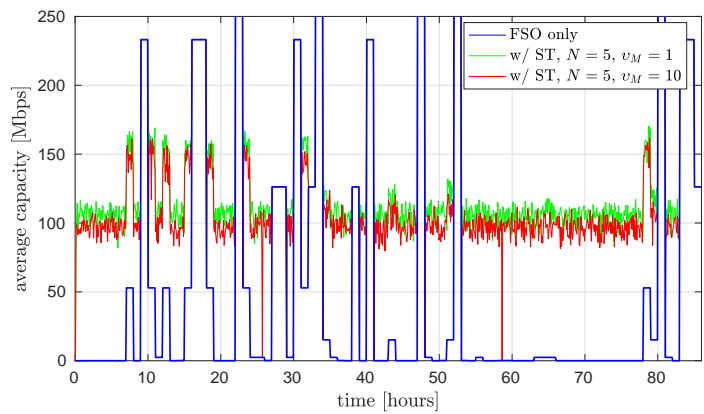}
	\caption{The average capacity versus the hours with fog in Edinburgh from January $2016$ to June $2017$. ST: spectrum trading.  }\label{F9_practical}
		\squeezeup
		\squeezeup
\end{figure}
In the simulation, we use the hours of fog events from the hourly visibility data (totally $\mathscr{H}_\mathrm{fog}=86$ hours) and assume that the coherence time of the weather conditions is $T_c=1$ hour, the time-scale of the traffic load change in RF nodes $T_t$ is $5$ minutes, and the coherence time of the RF fading is $T_f=3$ s. The weather-dependent coefficient $\kappa$ can be calculated based on the visibility using (\ref{visi}). Figure \ref{F9_practical} represents the simulation results and confirms that the FSO-only system is significantly outperformed by those systems with proposed spectrum trading between the source and RF nodes. It also shows that the system with relatively lower traffic loads in RF nodes ($\upsilon_M=1$) performs the best. If we consider the minimal data rate requirement $C_{th}=80$ Mbps as the threshold of the system below which an outage occurs, it can be calculated that during these fogy hours the outage probability for FSO-only system is $80\%$, whereas for the system with spectrum trading and $\upsilon_M=1$, the outage probability significantly reduces to $1.2\%$. 

{It is also valuable to evaluate the link availability enhancement of the proposed system over FSO-only link. The availability is defined as the probability of time in which the atmospheric loss does not cause outage and it is usually calculated over a period of one or more years \cite{Kim982}. Assuming $C_{th}=80$ Mbps, the availability of FSO-only link is given by $A_o=(\mathscr{H}_\mathrm{tot}-\mathscr{H}_\mathrm{out})/\mathscr{H}_\mathrm{tot}=99.4 \%$, where $\mathscr{H}_\mathrm{out}$ denotes the number of outage hours when $\overline{C}_{o}<C_{th}$. When the proposed system is employed, the link availability can be expressed as
\begin{equation}\label{ava}
\mathrm{availability}=\frac{\sum_{j=1}^{\mathscr{H}_\mathrm{out}}P_j^\mathrm{ava}+(\mathscr{H}_\mathrm{tot}-\mathscr{H}_\mathrm{out})\times 1}{\mathscr{H}_\mathrm{tot}},
\end{equation}               
where $P_j^\mathrm{ava}$ refers to the the availability probability of the proposed system at the $j$th outage hour experienced by the FSO-only system. Note that because of the randomness of the fading of RF link and traffic load in RF nodes, there might still remain a small probability of outage despite using the proposed system. Based on (\ref{ava}) and a Monte Carlo simulation, one can calculate that with $N=5$ an availability of $99.996 \%$ and $99.9997 \%$ can be achieved when $\upsilon_M=10$ and $\upsilon_M=1$, respectively. Therefore, the proposed system can significantly enhance the availability towards the five-nine carrier-class availability requirement.}

\section{Conclusion} \label{conclusion}
In this work, a novel hybrid RF/FSO system based on the market-equilibrium spectrum trading is proposed. It is assumed that no RF spectrum is preallocated to the FSO link, but when the FSO link availability is significantly impaired by the infrequent adverse weather conditions, the source can borrow a portion of licensed RF spectrum from one of the surrounding RF nodes to establish a dual-hop RF/FSO hybrid link to maintain its throughput to the destination. Through numerical simulations, it is shown that in adverse weather conditions the proposed scheme can significantly improve the average capacity of the system especially when the surrounding RF nodes are with low traffic loads. In addition, with the increase of the number of RF nodes involved in the system, better connectivity can be achieved by means of diversity, particularly when the RF node candidates have higher probability of being in heavy traffic load conditions. The substantial enhancement of average capacity and availability of the proposed system over FSO-only link is also demonstrated by applying it into a realistic channel based on the Edinburgh weather statistics.

	\bibliographystyle{IEEEtran}
	\bibliography{IEEEabrv,spectrum_trading}
\end{document}